\documentclass[sigconf, authorversion=true]{acmart}
\usepackage[utf8]{inputenc}
\usepackage{multirow}
\usepackage{natbib}
\usepackage{float}
\usepackage{appendix}
\usepackage{array}
\usepackage{hyperref}
\usepackage{graphicx}
\usepackage[
type={CC},
modifier={by-nc-sa},
version={3.0}
]{doclicense}
\PassOptionsToPackage{numbers, sort&compress}{natbib}
\PassOptionsToPackage{colorlinks=true, linkcolor=blue, filecolor=magenta, urlcolor=cyan}{hyperref}
\restylefloat{table}

\AtBeginDocument{%
  \providecommand\BibTeX{{%
    \normalfont B\kern-0.5em{\scshape i\kern-0.25em b}\kern-0.8em\TeX}
    }
}

\copyrightyear{2024}
\acmYear{2024}
\setcopyright{cc}
\setcctype[4.0]{by}
\acmConference[ASSETS '24]{The 26th International ACM SIGACCESS Conference on Computers and Accessibility}{October 27--30, 2024}{St. John's, NL, Canada}
\acmBooktitle{The 26th International ACM SIGACCESS Conference on Computers and Accessibility (ASSETS '24), October 27--30, 2024, St. John's, NL, Canada}
\acmDOI{10.1145/3663548.3675629}
\acmISBN{979-8-4007-0677-6/24/10}

\begin{document}

\title[EmoBridge: Bridging the Communication Gap between Students with Disabilities and Peer Note-Takers]{EmoBridge: Bridging the Communication Gap between Students with Disabilities and Peer Note-Takers Utilizing Emojis and Real-Time Sharing}

\author{Hyungwoo Song}
\authornote{These authors contributed equally to this research.}
\orcid{0009-0008-6649-9453}
\affiliation{%
    \department{Human Centered Computing Lab}
    \institution{Seoul National University}
    \city{Seoul}
    \country{Korea, Republic of}
}
\email{rotto95@snu.ac.kr}

\author{Minjeong Shin}
\authornotemark[1]
\orcid{0009-0003-1235-3436}
\affiliation{%
    \department{Information Science and Culture Studies}
    \institution{Seoul National University}
    \city{Seoul}
    \country{Korea, Republic of}
}
\email{shinmj1024@snu.ac.kr}

\author{Hyehyun Chu}
\authornotemark[1]
\orcid{0009-0006-0256-5277}
\affiliation{%
    \department{Information Science and Culture Studies}
  \institution{Seoul National University}
  \city{Seoul}
  \country{Korea, Republic of}
}
\email{ded06031@snu.ac.kr}

\author{Jiin Hong}
\authornotemark[1]
\orcid{0009-0009-8219-8252}
\affiliation{%
    \department{Information Science and Culture Studies}
  \institution{Seoul National University}
  \city{Seoul}
  \country{Korea, Republic of}
}
\email{snujiin@snu.ac.kr}

\author{Jaechan Lee}
\orcid{0009-0007-8149-8275}
\affiliation{%
  \department{Department of Computer Science and Engineering}
  \institution{Seoul National University}
  \city{Seoul}
  \country{Korea, Republic of}
}
\email{dlwocks31@snu.ac.kr}

\author{Jinsu Eun}
\orcid{0000-0003-3051-7193}
\affiliation{%
    \department{HCI+D Lab.}
  \institution{Seoul National University}
  \city{Seoul}
  \country{Korea, Republic of}
}
\email{eunjs71@snu.ac.kr}

\author{Hajin Lim}
\orcid{0000-0002-4746-2144}
\authornote{Corresponding author.}
\affiliation{%
  \department{Information Science and Culture Studies}
  \institution{Seoul National University}
  \city{Seoul}
  \country{Korea, Republic of}
}
\email{hajin@snu.ac.kr}

\renewcommand{\shortauthors}{Song, et al.}

\begin{abstract}
Students with disabilities (SWDs) often struggle with note-taking during lectures. Therefore, many higher education institutions have implemented peer note-taking programs (PNTPs), where peer note-takers (PNTs) assist SWDs in taking lecture notes. To better understand the experiences of SWDs and PNTs, we conducted semi-structured interviews with eight SWDs and eight PNTs. We found that the interaction between SWDs and PNTs was predominantly unidirectional, highlighting specific needs and challenges. In response, we developed EmoBridge, a collaborative note-taking platform that facilitates real-time collaboration and communication between PNT-SWD pairs using emojis. We evaluated EmoBridge through an in-the-wild study with seven PNT-SWD pairs. The results showed improved class participation for SWDs and a reduced sense of sole responsibility for PNTs. Based on these insights, we discuss design implications for collaborative note-taking systems aimed at enhancing PNTPs and fostering more effective and inclusive educational experiences for SWDs. 
\end{abstract}


\keywords{Students with disabilities, university disability support, student accessibility, peer note-taking program, peer notetaker, notetaking support, collaborative note-taking, higher education, emoji}



\maketitle

\begin{figure*}
  \includegraphics[width=\textwidth]{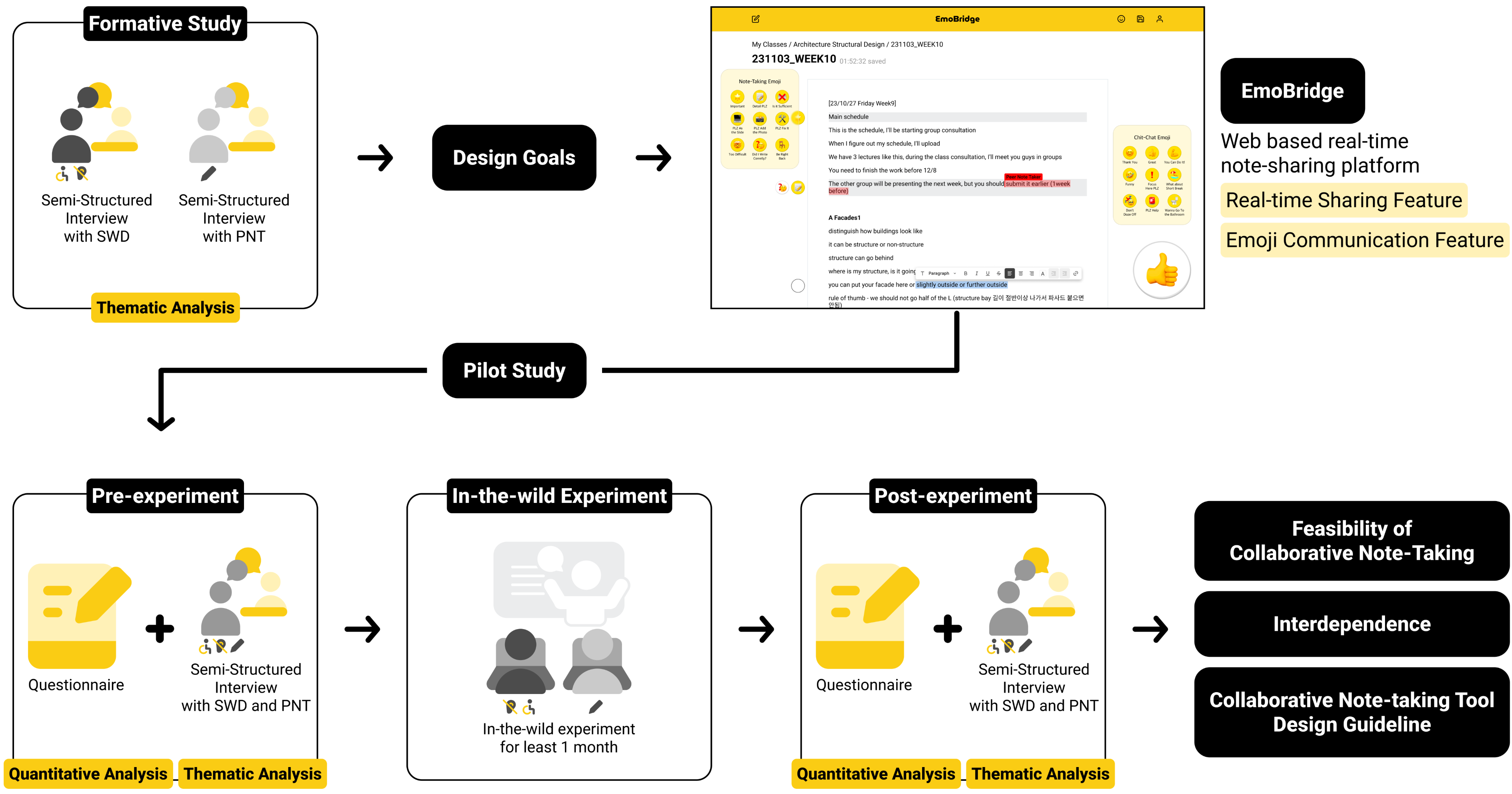}
  \caption{Research overview. This figure outlines the overall research procedure taken in this study. As a first step, semi-structured interviews were conducted with students with disabilities (SWDs) and peer note-takers (PNTs), highlighting their needs and challenges, particularly the unidirectional nature of their interaction. This informed the design of the collaborative note-taking platform, EmoBridge, a web-based note-taking platform facilitating real-time note-sharing and emoji-based communication. Following this, a pilot study was conducted to refine the design of EmoBridge, and an 'in-the-wild' evaluation was conducted with seven SWD-PNT pairs. The study concluded with design implications for improving PNTP and student accessibility and disability support in higher education institutions. }
  
  \Description{This figure illustrates the research and development process of EmoBridge. The process began with a formative study, where semi-structured interviews were conducted with students experiencing different types of disabilities (physical, hearing, and visual impairments) and with peer note-takers. This led to the formulation of specific design goals for EmoBridge. Using these goals, the EmoBridge platform was developed to enable real-time note-sharing and communication through emojis.
  Following the formative study, a pilot study was conducted, involving deployments and semi-structured interviews with SWDs (Students With Disabilities) and PNTs (Peer Note-Takers). This was followed by an in-the-wild deployment and a post-deployment analysis that included thematic and quantitative analyses of questionnaires, contributing to the overall development of design guidelines for future enhancements to EmoBridge.}
  \label{fig:teaser}
\end{figure*}

\section{Introduction}

In higher education settings, students with disabilities (SWDs) encounter numerous obstacles, particularly in attending classes \cite{barriers, schoolproblems}. SWDs often face challenges in keeping up with lectures and taking notes, which can negatively impact their academic performance and outcomes \cite{murray2000postsecondary}. Therefore, many higher education institutions have implemented Peer Note-Taking Programs (PNTPs) that pair SWDs with peers in class (e.g., \cite{cam_notetaking, cmu_notetaking, cornell_notetaking}). PNTPs usually involve peer note-takers (PNTs) who assist SWDs with note-taking, either by transcribing notes during class or by providing their own notes after class \cite{umn_notetaking, usc_notetaking}. 

While PNTPs provide some support to SWDs, they often fall short in promoting active participation from SWDs. Prior studies showed that the interaction between SWDs and PNTs is typically unidirectional, with PNTs providing support without receiving feedback or engagement from SWDs \cite{ChoSon2013}. However, research on solutions to address these issues and on the specific needs of each participant remains limited \cite{Kim2011}.

Our study aimed to address these gaps by deeply understanding the interactions and specific needs of PNTP participants. We also explored the possibility of applying collaborative note-taking to PNTP, which is known to enhance class participation and learning efficiency \cite{harbin2020collaborative}. In the end, we proposed guidelines for collaborative note-taking systems aimed at improving PNTPs, so as to promote more inclusive and effective educational experiences for SWDs. 

To understand the specific needs and challenges associated with PNTPs, we conducted a formative study using semi-structured interviews with 16 individuals (8 SWDs and 8 PNTs) experienced in PNTPs. Our findings confirmed previous research highlighting that interactions in PNTPs remained predominantly unidirectional \cite{ChoSon2013}. Additionally, we discovered that SWDs desired to participate in note-taking during class, while PNTs wanted to receive feedback on their notes from the SWDs they supported. Based on these insights, we established design requirements to meet their needs.

Based on these design requirements, we designed and developed EmoBridge, a collaborative note-taking tool with real-time note-sharing and emoji communication features. After a pilot study to refine EmoBridge's design, we conducted an in-the-wild evaluation study with seven SWD-PNT pairs who were asked to utilize EmoBridge for PNTP over one month. Our findings revealed that SWDs felt more engaged in classes, experiencing increased self-efficacy as they contributed to note-taking. Concurrently, PNTs found they could rely more on SWDs, which reduced some of their psychological burden of sole note-taking responsibility. Furthermore, we discovered that diverse SWD-PNT pair relationship dynamics and class contexts significantly influenced EmoBridge's usage patterns and satisfaction levels across pairs. Factors such as the closeness of relationships between SWDs and PNTs, as well as class-specific elements like instructional style, played crucial roles in shaping user experiences with the EmoBridge.

Our findings indicate that integrating collaborative note-taking platforms like EmoBridge into PNTPs could significantly enhance their effectiveness. Drawing from these insights, we proposed design implications for collaborative note-taking tools that not only facilitate effective note-taking but also foster a sense of ``interdependence'' between SWDs and PNTs within PNTPs, improving disability support and student accessibility in higher educations. An overview of our research process is shown in Fig. \ref{fig:teaser}.

The contributions of this study are as follows:
\begin{itemize}
    \item Identifying specific needs and challenges arising from unidirectional interactions between SWDs and PNTs within PNTPs through a semi-structured interview study.
    \item Designing and implementing EmoBridge, a web-based collaborative note-taking platform for supporting PNTP.
    \item Evaluating EmoBridge in real classroom environments by evaluating seven pairs of SWD and PNT in the field.
    \item Proposing guidelines for collaborative note-taking tools to improve the efficacy of PNTPs and enhance student accessibility and disability support.
\end{itemize}

\section{Related Work}

\subsection{Challenges and Supports for Students with Disabilities in Higher Education}

In higher education, students with disabilities (SWDs) face numerous challenges \cite{barriers, schoolproblems}. Many university and college facilities include physical barriers that make it difficult for SWDs to navigate \cite{law2007perceived}, often hindering their active engagement in social activities \cite{hemmingson2002environmental}. Additionally, research indicates that SWDs encounter substantial academic challenges \cite{murray2000postsecondary}. These challenges include limited access to course materials \cite{coursematerial} and difficulties in note-taking, which limit participation in classes and impede learning \cite{hughes1994note}.

To address these challenges, federal laws such as the Americans with Disabilities Act (ADA) \cite{americans} and the Individuals with Disabilities Education Act (IDEA) \cite{individuals} mandate that SWDs must be provided with accommodations to ensure access to educational opportunities and resources. As a result, centers for students with disabilities at higher education institutions offer a range of programs designed to assist SWDs academically \cite{stodden2001current, CSD1, CSD2, CSD3}. These programs include offering educational resources and operating support programs tailored to the specific needs of each SWD \cite{washington}. Such need for these personalized supports arises from the varying degrees of disability among students, necessitating personalized assistance \cite{hadley2011college}. One form of personalized assistance is personnel support, which may involve assistance from specially trained professionals \cite{dundee2024nonmedical} or peer students \cite{CMU}.

\subsection{Peer Note-Taking Program}

As part of the personnel support provided by centers for students with disabilities at higher education institutions, the Peer Note-Taking Program (PNTP) is commonly adopted (e.g., \cite{cam_notetaking, cmu_notetaking, qub_notetaking, umn_notetaking, snu_notetaking, ung_notetaking, usc_notetaking, cornell_notetaking, u-paris_notetaking, gatech_notetaking, louisiana_notetaking}). In many cases, PNTPs involve peer supporters, known as peer note-takers (PNTs), who assist SWDs with note-taking by either transcribing notes during class or providing their own notes after the class. 

The operational process of PNTPs varies across institutions. For instance, some universities administer the PNTP via online platforms, enabling PNTs to share their notes with SWDs asynchronously (e.g., \cite{cmu_notetaking, cornell_notetaking}). In other institutions, PNTs are instructed to provide direct assistance to SWDs during lectures (e.g., \cite{rit2024disabilityservices}).

Despite the prevalence of PNTPs, research on their effectiveness remains limited. Existing studies have highlighted that interaction between SWDs and PNTs is predominantly unidirectional, typically flowing from PNTs to SWDs \cite{ChoSon2013, Kim2011}. As a result, SWDs often remain passive recipients of support, and PNTs tend to feel less satisfied with PNTP programs due to the lack of interaction with SWDs \cite{Choi2019}. As such, current research focuses primarily on the challenges facing either SWDs or PNTs within PNTPs, but has not investigated viable strategies or solutions to improve the unidirectional dynamics of their interactions.

Therefore, this study aims to gain a deeper understanding of the interactions between SWDs and PNTs within classroom settings and to gain insights that could help address the issues associated with unidirectional communication and enhance the effectiveness of PNTPs. To achieve this, we conducted a formative study with eight SWDs and eight PNTs.

\subsection{Collaborative Note-Taking}

Collaborative note-taking involves two or more peers collectively taking notes, annotating, exchanging feedback, or asking questions about the content being learned \cite{costley2021collaborative}. Extensive research has documented its benefits \cite{harbin2020collaborative, singh2004collaborative}, including promoting student cooperation, increasing engagement and concentration during lectures \cite{zyto_successful_2012}, and enhancing learning efficiency \cite{luo_revising_2016,fanguy_analyzing_2023, courtney2022individual}. Further, a large body of research has explored the use of digital technologies to facilitate collaborative note-taking, as a form of mobile apps \cite{smartphone,popescu2016edunotes} or web-based tools \cite{zyto_successful_2012, petko2019metapholio}.

In fact, PNTP creates an ideal environment for applying collaborative note-taking as it inherently involves the participation of both PNTs and SWDs. However, the current practice and facilitation of PNTPs often result in predominantly unidirectional communication, leaving SWDs in a passive role during the note-taking process and limiting the benefits of collaboration. Therefore, the primary goal of this study is to explore the potential of integrating collaborative note-taking into PNTPs. By developing and evaluating EmoBridge through in-the-wild evaluation studies involving seven SWD-PNT pairs, we investigated the effects and consideration of integrating collaborative note-taking into real-world PNTPs.

\section{Formative Study}

To explore the needs and challenges faced by PNTs and SWDs in class settings, we conducted a formative study using semi-structured interviews. By understanding the needs of each group, we established design requirements to integrate collaborative note-taking into PNTP practices.

\subsection{Method}
A total of 16 participants were recruited for formative interviews, consisting of eight SWDs (four with mobility impairments and four with hearing impairments) and eight PNTs (two who worked with hearing-impaired students, two with visually impaired students, one with a mobility-impaired student, and three with students having multiple impairments). Recruitment of participants was conducted in collaboration with the Centers for Students with Disabilities at multiple universities. All of our research procedures were approved by the Institutional Review Board (IRB) of Seoul National University. 

Recruitment was facilitated through 14 university Centers for Students with Disabilities. Contacts were made with SWDs and PNTs at each center, and pre-surveys were distributed via email to registered individuals. These pre-surveys included brief questions about their experiences with PNTP and their contact information. Only those who volunteered and had at least one semester of PNTP experience were eligible to participate. Thus, despite contacting 14 universities, participants were recruited only from 8 universities. The study goal and procedure were thoroughly explained to those who were eligible, and written informed consent was obtained from each participant.

\begin{table}[t]
\centering
\resizebox{\columnwidth}{!}{
\begin{tabular}{lllll}
\toprule
ID & Role & Type& Gender & University \\ 
\midrule
SWD-1 & SWD & Physical & F & U1 \\
SWD-2 & SWD & Physical & M & U5 \\
SWD-3 & SWD & Hearing & M & U6 \\
SWD-4 & SWD & Hearing & F & U8 \\
SWD-5 & SWD & Hearing & M & U6 \\
SWD-6 & SWD & Physical & M & U2 \\
SWD-7 & SWD & Physical & M & U1 \\
SWD-8 & SWD & Hearing & M & U2 \\
\midrule
PNT-1 & PNT & Physical & F & U1 \\
PNT-2 & PNT & Hearing & F & U1 \\
PNT-3 & PNT & Hearing & M & U4 \\
PNT-4 & PNT & Vision & F & U3 \\
PNT-5 & PNT & Vision & F & U1 \\
PNT-6 & PNT & Physical, Hearing & F & U1 \\
PNT-7 & PNT & Hearing, Vision & F & U1 \\
PNT-8 & PNT & Hearing, Vision & F & U2 \\
\bottomrule
\end{tabular}
}
\caption{Formative Study Participants. The (ID) column denotes the ID assigned to each participant. The (Role) column indicates the roles participants had undertaken within PNTP. The (Type) column specifies the type of disability for SWDs and the type of disability of SWDs supported by each PNT participant. The (Gender) column denotes the gender of participants (F: female; M: male). The (University) column indicates the ID of the affiliated universities of the participants.}

\Description{This table provides detailed information about the Formative Study Participants, including their ID, Role, Type, Gender, and University. The (ID) column assigns an index to each participant. The (Role) column describes the roles participants undertook in the PNTP. The (Type) column differentiates between the types of disabilities for SWDs and the types of disabilities supported by PNTs. The (Gender) column specifies the gender of participants, with M for Male and F for Female. The (University) column lists the universities affiliated with the participants.}
\label{table:formativestudyparticipants}
\end{table}

The interviews were conducted either face-to-face or remotely via Zoom, with each session lasting about an hour. Consent for audio recording was obtained from all participants before the sessions, and the interviews were subsequently recorded. In each interview, one member of the research team led the session as the primary interviewer, while the other took detailed field notes. As part of the interview questions, participants were invited to share their lived experiences with PNTP. For example, we asked SWDs about their specific disabilities, the type of assistance they required, and their expectations of PNTs. Meanwhile, PNTs were questioned about their motivations for joining PNTP, their individual note-taking styles, the approaches they worked as PNTs, and any suggestions they had for enhancing the program.

In analyzing the transcripts, we employed a thematic analysis  \cite{saldana2021coding}. Initially, the four members of the research team independently conducted open coding. Following this, a review round was held where researchers examined and compared each other’s codes. Subsequently, the research team collectively engaged in several rounds of discussion to refine these initial codes into broader and overreaching themes, such as `in-class dynamics and feedback,' and ‘challenges and burdens.’ As part of this iterative process, the research team members collaboratively worked on developing the themes and establishing the design goals while resolving disagreements through discussion.


\subsection{Findings}
The findings of our formative study corresponded with those of earlier studies \cite{ChoSon2013, Kim2011} that communication between SWDs and PNTs was predominantly unidirectional. In most cases, PNTs took notes on their local document writing programs and shared them with SWDs after class. SWD-8 mentioned, ``\textit{When my PNT shared notes, she just uploaded them to a shared (google) drive after class. So, the actual note-taking support is only available after class. There is no real-time interaction between me and my PNT}.''

We found that the classroom environment further constrained communication between SWDs and PNTs. Both parties were cautious about giving feedback during class, as they did not want to disturb each other or the other students. PNT-2 described, ``\textit{Since we need to remain quiet during class and the pace of instruction is very fast, I type at around 700 characters per minute, leaving me no time to rest or check on the surroundings (including SWD)}.''

Due to this unidirectional interaction constrained by classroom settings, minimal interaction and lack of feedback between SWDs and PNTs were prevalent. Consequently, many SWDs and PNTs faced various challenges and exhibited a need for more active interaction and feedback.

However, PNT participants working with visually impaired SWDs reported distinct in-class experiences for these students compared to other SWDs. Visually impaired students relied heavily on auditory channels for information acquisition and found it challenging to process multiple sources of information beyond the lecturer's voice during classes. Participant PNT-4 noted, ``\textit{My (visually impaired) peer primarily communicates through voice. Beyond listening to the lesson, reading materials or taking notes is challenging as using a screen reader makes it difficult to follow the lesson}.''
Given that visually impaired students seem to require a different approach to in-class interaction regarding PNTP compared to students with other types of disabilities, the remainder of this paper will focus on SWDs with vision and their corresponding PNTs. Below, we outlined the main findings from the formative study. 

\paragraph{\textbf{Finding \#1: SWD participants desired real-time note sharing during class.}} The majority of SWD participants encountered difficulties in monitoring the notes taken by PNTs during class, despite their desire to do so. They explained that they needed two to three times more time to fully comprehend notes that were not tracked in real-time during the class. SWD-7 stated, ``\textit{Since my PNTs usually sat next to me, I tried to see their notes by glancing at their laptop screen from the side, but it was not easy to see}.''

\paragraph{\textbf{Finding \#2: SWD participants wanted more involvement in PNT's note-taking during class.}} In addition, some SWD participants wanted more involvement in note-taking in order to prevent their PNT from recording irrelevant information or missing lecture content. SWD-1 said, ``\textit{Even if my PNT takes diligent notes, they might lack detail on important points. Therefore, I usually record the lecture and listen to my own recordings during the exam periods}.'' As such, despite the fact that PNTs' notes were sometimes incomplete and unsatisfactory, all SWDs preferred to correct the notes directly rather than discuss them with the PNT out of consideration for their PNTs.

\paragraph{\textbf{Finding \#3: PNT participants wanted feedback on their notes from SWDs.}} While many SWDs did not comment on PNT's notes, PNT participants expressed a strong desire for feedback on their note-taking. They wanted direct input from SWDs on the quality and usefulness of their notes. PNT-3 stated, ``\textit{I wanted to get some feedback on my notes from my SWD peer, but it was technically challenging. I hope there are ways to facilitate active communication with SWD and determine good note-taking methods that satisfy both parties}.'' 

\paragraph{\textbf{Finding \#4: PNT participants experienced psychological burden from sole responsibility of note-taking.}} Many PNT participants reported feeling overburdened by being solely responsible for note-taking. This burden intensified when handling unfamiliar class content. PNT-7 stated, ``\textit{I felt pressured not to miss any class content. Unlike my own classes, I reviewed notes thoroughly multiple times. [...] Law classes were the most challenging due to the difficult legal terminology and my lack of prior knowledge}.'' Consequently, PNT participants said they tried to take meticulous notes due to this pressure. However, they often felt stressed due to the lack of feedback or ways to know whether their notes were helpful for SWDs.

\paragraph{\textbf{Finding \#5: Both SWD and PNT participants recognized the need for collaborative partnership}}
While the level of intimacy between SWD-PNT pairs varied, all participants acknowledged the necessity of a collaborative relationship. They saw some level of collaborative partnership would be more beneficial than current unidirectional relationships for PNTP. They also believed that such relationship dynamics would facilitate discussions about note-taking and the exchange of feedback. SWD-6 emphasized this point, stating, ``\textit{Having a suitable level of rapport would be essential for providing effective feedback related to note-taking}.''

\subsection{Design Requirements}
As highlighted in the Findings, the interaction and communication between SWDs and PNTs was largely unidirectional. SWD participants desired a more active role in in-class note-taking, including real-time tracking and intervention, but were hesitant to do so out of consideration for their PNTs. Similarly, PNTs wanted input and feedback from SWDs to ensure their notes were helpful, but experienced psychological burden from the sole responsibility of note-taking.

To address these challenges, we envision shared note-taking platforms that could facilitate in-class interaction between PNTs and SWDs. Such platforms would be beneficial in addressing the identified challenges and needs of both parties and fostering collaborative partnerships between SWDs and PNTs. Based on the findings and these considerations, we identified the following design requirements for a collaborative note-taking system for PNTP:

\paragraph{\textbf{Requirement \#1: Provide opportunities for SWDs to contribute to note-taking} (Findings \#1, 2)}
A collaborative note-taking platform for PNTP would be necessary to ensure that SWD could contribute to note-taking, starting with accessing notes during class, tracking them, and providing feedback to PNTs. The real-time sharing of notes will satisfy SWDs' need to monitor notes, as well as serve as a foundation for SWDs to actively participate in note-taking. 

\paragraph{\textbf{Requirement \#2: Alleviate the mental burden of PNTs}  (Findings \#3, 4)} 
The mental burden PNTs experienced stemmed from their perception of sole responsibility for note-taking. Their psychological stress was exacerbated by the lack of means to determine if their note-taking methods were effective and if their notes were useful for SWDs. Therefore, a collaborative note-taking system for PNTP will need to aim to alleviate this burden by fostering shared responsibility for note-taking by involving SWDs in the process. In doing so, enabling PNTs to receive feedback or positive affirmation on their notes from SWDs would be beneficial. These would help distribute the responsibility and provide PNTs with the reassurance they need about the value and effectiveness of their work.

\paragraph{\textbf{Requirement \#3: Facilitate suitable in-class communication between SWDs and PNTs}} 
Sharing feedback on note-taking between SWDs and PNTs would be beneficial. However, in the formative study, there was a need to avoid disrupting the class. Therefore, communication methods should not disrupt the class. For example, quiet and non-distracting communication methods would be beneficial to ensure that the lecturer and other students would not be disturbed. It would also be important that these methods do not disrupt the users' concentration while following the lecture.

\paragraph{\textbf{Requirement \#4: Facilitate Collaborative Partnership} (Finding \#5)} 
A collaborative note-taking tool for PNTP will need to serve as a means for PNTs and SWDs to build a collaborative partnership. To this end, it would be crucial to promote mutual recognition of note-taking as a shared goal and support collaborative efforts toward achieving it. By fostering this collaborative partnership, SWDs and PNTs could both enhance the overall effectiveness of the note-taking process and nurture a sense of shared responsibility and mutual support in PNTPs.

\begin{figure*}[h!]
  \centering
  \includegraphics[width=\textwidth]{Figure2_EmoBridge.png}
  \caption{EmoBridge Web Interface. (a) Toggle switch for the note-taking emoji window. (b) Field for the note title. (c) Note-taking emoji window. (d) Text highlighted in gray where a note-taking emoji is inserted. (e) Inserted note-taking emoji. (f) Real-time mutual cursor of the partner. (g) Indicator showing the selected text line for note-taking emoji insertion. (h) Text format menu. (i) Inserted chit-chat emoji. (j) Chit-chat emoji window. (k) Toggle switch for the chit-chat emoji window. (l) Note-saving button for exporting as a txt file. (m) Log-in/out button.}
  
  \Description{EmoBridge Web Interface. This figure illustrates various elements of the EmoBridge note-taking page: (a) Toggle switch to open/close the note-taking emoji window, (b) Note title field, (c) Note-taking emoji window with different emojis for feedback, (d) Text highlighted in gray indicating where a note-taking emoji is inserted, (e) Inserted note-taking emoji, (f) Real-time mutual cursor position of a collaborative partner, (g) Selected text line indicator for note-taking emoji insertion, (h) Text format menu for text adjustments, (i) Inserted chit-chat emoji, (j) Chit-chat emoji window for general communication, (k) Toggle switch for the chit-chat emoji window, (l) Button to save and export notes as a txt file, and (m) Log-in/out button. This diagram provides an overview of the key features and user interactions on the EmoBridge note-taking page, highlighting the platform’s functionality and tools.}
  \label{figure:EmoBridge_Interface}
\end{figure*}

\section{EmoBridge}

EmoBridge is a collaborative note-taking and sharing platform specifically designed for PNTPs. It allows both SWDs and PNTs to view the note-taking screen on their respective devices and exchange feedback on notes via emojis, as shown in Fig. \ref{figure
}. In addition to facilitating communication about note-taking, EmoBridge enables users to share their emotions and states experienced during class time through emojis.

\subsection{Features}

EmoBridge's key functionalities were designed to align with the design requirements identified in Section 3.3. First, the real-time editing and sharing of the note-taking screen enable SWDs to directly engage and contribute to note-taking (\textbf{Requirement \#1}). This allows both SWDs and PNTs to view the same notes on their respective devices during class and provide feedback to each other, thereby reducing the burden on PNTs through SWD's feedback or direct edits (\textbf{Requirement \#2}). Secondly, the emoji feature was designed to foster and maintain a collaborative partnership between SWDs and PNTs \textbf{(Requirement \#4}). To support this, emojis were categorized into note-taking emojis and chit-chat emojis. This lightweight but convenient communication method, enabled by emojis, was devised with consideration of the class environment, ensuring it is neither demanding nor disruptive to other students or the users themselves (\textbf{Requirement \#3}).

In particular, \textbf{real-time editing and synchronization of note content} ensures that any content entered in the note-taking window is instantly shown across all devices of SWDs and PNTs in real-time. This feature allows SWDs to view the notes as they are created, enabling them to immediately identify and suggest changes or improvements during class. In addition, EmoBridge supports \textbf{mutual cursor visibility} that displays the position of each other’s cursor in the note-taking window in real-time. This allows both parties to see where the other is focusing within the note. 

\begin{figure}[t!]
  \centering
  \includegraphics[width=\linewidth]{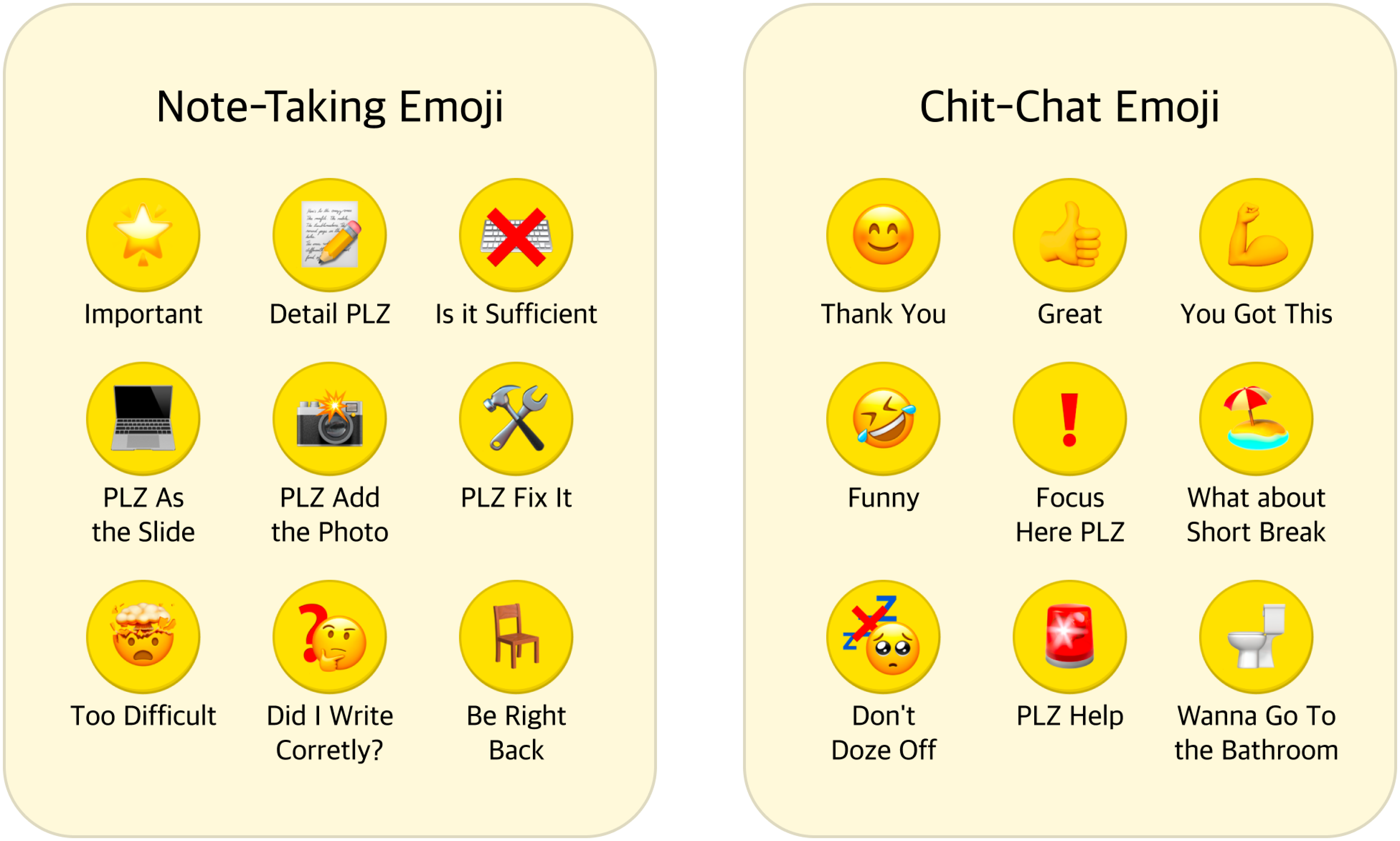}
  \caption{Emoji list of EmoBridge. Note-taking emoji on the left side, Chit-chat emoji on the right side.}
  
  \Description{The Emoji list of EmoBridge. This figure displays the available emojis in the EmoBridge platform, with note-taking emojis on the left and chit-chat emojis on the right. The note-taking Emojis include a variety of icons to convey feedback related to note-taking, such as "Important," "Detail PLZ," "Is Sufficient," "PLZ As the Slide," "PLZ Add the Photo," "PLZ Fix It," "Too Difficult," "Did I Write Correctly?", and "BRB (be right back)s." The Chit-chat Emojis are used for more general communication, containing emojis like "Thank You," "Great," "You Got This," "Funny," "Focus Here PLZ," "What about Short Break," "Don't Doze Off," "PLZ Help," and "Wanna Go To the Bathroom."}
  \label{figure:Emoji_List}
\end{figure}

In addition, to support in-class interaction between SWDs and PNTs, EmoBridge provides two sets of emojis that facilitate seamless exchanges of feedback about notes (\textbf{Note-Taking emojis}) and sharing of emotions and personal states experienced during class (\textbf{Chit-chat emojis}). The choice of emojis as a communication method is based on their ease and speed of use, allowing both SWDs and PNTs to communicate without much attention or effort, thereby not distracting from the lecture.

\textbf{Note-Taking Emojis} are designed for exchanging feedback about notes during class. Note-Taking emojis can be used for various purposes such as expressing importance (‘Important’), requesting elaboration (‘Detail PLZ’, ‘PLZ As in the slide’, ‘PLZ add the photo’), or asking for feedback on their own notes (‘Did I write correctly?’) as shown in Fig. \ref{figure:Emoji_List}. 

\textbf{Chit-chat Emojis} are intended to convey messages related to emotions or personal states. These include supportive emojis like 'Thank You', 'Great', 'You got this', and context or mood-sharing emojis such as 'Funny', 'PLZ focus here', 'What about short break', 'Don't doze off', 'PLZ help', 'Wanna go to the bathroom', also shown in Fig. \ref{figure:Emoji_List}. By facilitating expressions of gratitude and light communication, these emojis aim to build a collaborative partnership, fostering the development of a supportive and intimate communication layer between SWDs and PNTs.

\subsection{Implementation}
The implementation of EmoBridge focused on two main aspects: (1) facilitating real-time note and emoji synchronization, and (2) ensuring the management of note content through real-time backup. The EmoBridge website is accessible at \url{https://emobridge.vercel.app/}, and the source code of the EmoBridge is available at \url{https://github.com/dlwocks31/emobridge}.

The front-end development of EmoBridge was implemented using Next.js, along with Typescript, HTML, and CSS. For the web-based note editing functionality, EmoBridge utilized the BlockNote, a WYSIWYG (What You See Is What You Get) open-source editor \cite{blocknotejs}. We customized it to implement emoji input functionalities.

On the back-end development, notes were saved to the server in real-time through web sockets, ensuring instant content backup and minimizing data loss during unexpected shutdowns. The platform leveraged PartyKit and Cloudflare Durable Objects to enable real-time document editing and emoji synchronization with minimal latency \cite{partykit, cloudflare}. Supabase was employed for storing and managing user and document data, providing an SDK for efficient data management and a Google login system for secure and convenient user authorization \cite{supabase}.

\begin{figure*}[t!]
  \centering
  \includegraphics[width=\textwidth]{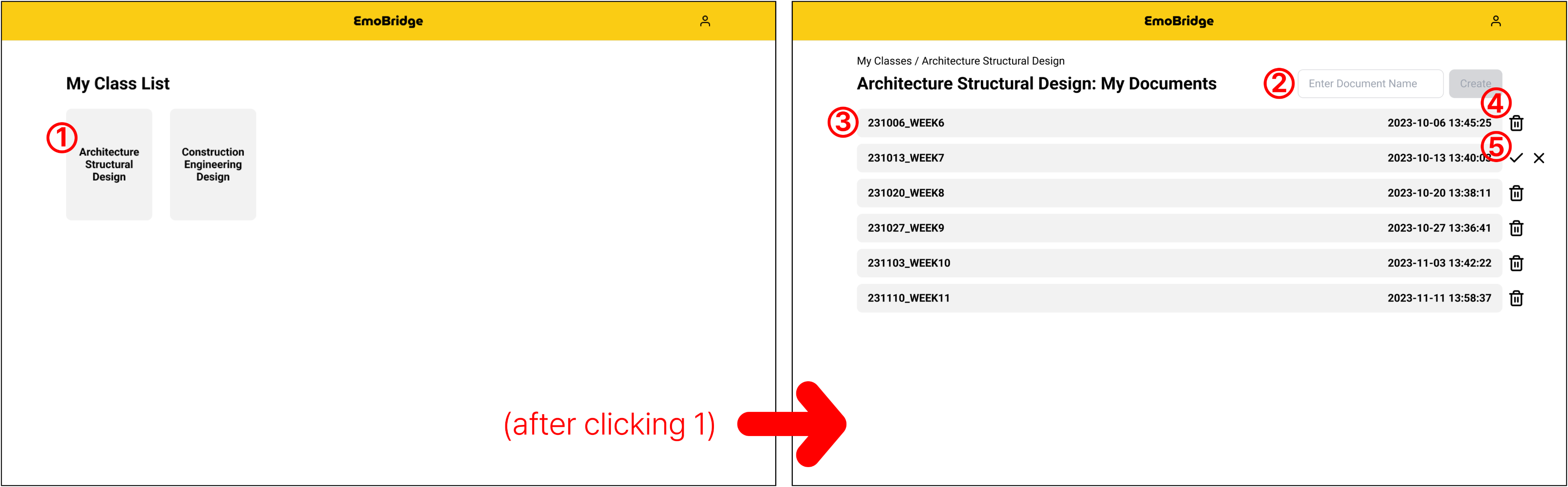}
  \caption{EmoBridge's note management folder system interface. When a user logs in, they are directed to the 'My Class' screen on the left, where they can view a list of their classes. By selecting (1) a class name, a user is taken to the 'My Documents' screen on the right, which displays a list of their notes. To create a document, (2) the user enters the note title in the top right corner and clicks the create button. To delete a document, they click the (3) trashcan button next to the document. Confirming the action with a (4) check mark will delete the note, while clicking the X button will cancel the deletion.}
  
  \Description{The interface for EmoBridge's note management and folder system. Upon login, users arrive at the ‘My Class’ interface, shown on the left, presenting a list of their courses. Clicking (1) on a course title leads to the ‘My Documents’ interface, depicted on the right, where users can see a catalogue of their notes. For new document creation, one inputs a title in the (2) entry field and initiates the process via the creation button. Documents are removable by selecting the (3) trashcan icon; confirmation is sought with a (4) check mark for deletion or the X button to abort.}
  \label{figure:Document_System}
\end{figure*}

\subsection{Example Use Scenario}
This section demonstrates an example user scenario of EmoBridge to illustrate how in-class interaction and collaborative note-taking can take place.

Before the class begins, both the SWD and PNT users open the EmoBridge webpage and log in. They are then directed to the ```My Class'' page, where they select the class name they are taking together. After the PNT user creates a new document for today's class, the SWD user joins the same page (see Fig. \ref{figure:Document_System}).

During the lecture, the PNT user takes notes, which the SWD user can view in real-time. At one point, the SWD notices that the PNT missed some detailed information mentioned by the instructor. To convey the need for more details, the SWD inserts the ``Detail PLZ'' emoji from the note-taking emoji window. To do this, the SWD opens the note-taking emoji window by clicking the top left Note-taking emoji icon of the navigation bar (see (a) in Fig. \ref{figure:EmoBridge_Interface}). The emoji window can be moved to anywhere on the page by dragging or closed by re-clicking the button on the navigation bar to avoid blocking the notes. Next, the SWD selects the relevant text block and inserts the ``Detail PLZ'' emoji (see (e) in Fig. \ref{figure:EmoBridge_Interface}). The text block where an emoji is inserted is highlighted in gray.  

Once the ``Detail PLZ'' emoji is inserted, the PNT notices it, moves the cursor to the corresponding text block, and adds more details about that section. After updating the notes, the PNT clicks the ``Detail PLZ'' emoji to indicate that the issue has been resolved.

Feeling grateful after seeing the updated notes, the SWD wants to express thanks by inserting a chit-chat emoji. To do this, the SWD clicks on the circular input area at the bottom right of the screen\footnote{This can also be done by clicking the Chit-chat emoji icon (see (k) in Fig. \ref{figure:EmoBridge_Interface})}, opening the chit-chat emoji window. This window can also be moved or closed, similarly to the note-taking emoji window. The SWD selects the ``Thank You'' emoji from the chit-chat emoji window. The inserted chit-chat emoji appears on the bottom right of both users' EmoBridge screens for 5 seconds before automatically disappearing (see (i) in Fig. \ref{figure:EmoBridge_Interface}). After seeing the ``Thank You'' emoji, the PNT responds by following the same process to insert a ``Great'' chit-chat emoji.

\subsection{Pilot Study}
Throughout the development process, we conducted a series of pilot studies to gain early insights into the usability of the EmoBridge interfaces. A total of three SWD-PNT pairs participated in the pilot study. They tried out the prototypes of EmoBridge and provided design feedback. Based on their feedback, key improvements were made to the interface of EmoBridge. For example, we enabled the emoji window to move freely, as the fixed location of the emoji window made the note-taking area narrower, and some SWDs with motor impairments found it hard to locate. Additionally, we added colored blocks with emojis in gray based on feedback that the emojis might be hard to notice during class. 
Through this iterative process of refining EmoBridge's interface based on user feedback, we developed the version of EmoBridge used in the following in-the-wild evaluation study for deployment in an actual classroom setting.

\section{In-the-wild evaluation of EmoBridge}
To evaluate the effectiveness of EmoBridge and examine the potential of integrating collaborative note-taking into PNTPs, we conducted in-the-wild evaluation studies involving seven SWD-PNT pairs who were actual participants of PNTP in the fall semester of 2023. Each pair of participants was asked to exclusively use the EmoBridge for note-taking during their classes. By identifying usage patterns and conducting both quantitative and qualitative evaluations of their experiences, we gained deeper insights into how EmoBridge influenced the participants’ PNTP experiences.

\subsection{Participants}

\begin{table}[ht!]
\centering
\resizebox{\columnwidth}{!}{
\begin{tabular}{llllll}
\toprule
ID &  Role & Type & Gender & University \\
\midrule
SWD-1  & SWD & Physical & F & U1 \\
SWD-2 & SWD & Physical & M & U5 \\
SWD-3 & SWD & Hearing & M & U6 \\
SWD-4 & SWD & Hearing & F & U8 \\
SWD-5 & SWD & Hearing & M & U6 \\
\midrule
PNT-9  & PNT & Physical & F & U1 \\
PNT-10  & PNT & Physical & F & U5 \\
PNT-11 & PNT & Hearing & F & U6 \\
PNT-12  & PNT & Hearing & M & U8 \\
PNT-13  & PNT & Hearing & F & U8 \\
PNT-14  & PNT & Hearing & M & U6 \\
PNT-15  & PNT & Hearing & F & U6 \\
\bottomrule
\end{tabular}
}
\caption{In-the-wild Evaluation Study Participants. The (ID) column indicates the ID assigned to participants. The (Role) column describes the roles undertaken by participants in PNTP. The (Type) column specifies the types of disabilities for SWD and the types of disabilities of SWDs supported by PNTs. The (Gender) column indicates the gender of participants (F: female; M: male). The (University) column indicates the ID of the affiliated universities of the participants.}

\Description{This table presents detailed information about the Evaluation Study Participants, including their ID, Formative Study ID, Role, Type, Gender, and University. The (ID) column indicates the index assigned to participants. The (Role) column lists the roles undertaken by participants in PNTP. The (Type) column describes the types of disabilities for SWD and the types of disabilities of SWD supported by PNT. The (Gender) column indicates the gender of participants, with M representing Male and F representing Female. The (University) column identifies the participants’ affiliated universities.}
\label{table:evaluationstudyparticipants}
\end{table}

A total of seven SWD-PNT pairs participating in the PNTP in the fall semester of 2023 took part in the study, including five SWDs and seven PNTs. The unequal numbers of SWDs and PNTs were due to two SWDs each participating with two PNTs. Similar to the formative study, we contacted the Centers for Students with Disabilities at multiple universities and attempted to recruit participants. However, recruitment of new SWD participants was not possible, so we reached out to previous SWD participants from the formative study for additional involvement. As a result, all SWD participants from the formative study agreed to participate in the evaluation study, while all PNTs who were paired with SWD participants were newly recruited. Detailed information about participants in the evaluation study can be found in Table \ref{table:evaluationstudyparticipants}.

\subsection{Study Procedure}
The evaluation study procedure was divided into three phases: pre-deployment, deployment, and post-deployment phases.

\subsubsection{Pre-deployment Phase} The study commenced one month after the start of the fall semester of 2023, providing participants who were newly acquainted with PNTP and new partners the opportunity to initially engage in their usual note-taking practices. After obtaining the study consent, participants were asked to complete a questionnaire to evaluate diverse aspects of their PNTP experiences during the initial month of classes. The questionnaire items are listed in Table \ref{table:questionnaire}. This questionnaire was completed again after the deployment phase to evaluate changes in their PNTP experiences after using EmoBridge for at least four weeks.

Subsequently, participants participated in a 30-minute semi-structured interview. The interviews were designed to allow participants to freely share their initial PNTP experiences and provide in-depth explanations for their responses to each questionnaire item. In the interviews, one researcher led the conversation while another researcher captured detailed notes and posed additional questions. Following the interviews, each participant received a tutorial on using EmoBridge. This pre-deployment phase procedure was held either in person or via online platforms and took about one hour.

\subsubsection{Deployment Phase}
During the deployment phase, all SWD-PNT pairs were required to utilize EmoBridge exclusively for note-taking in class for a minimum of four weeks. The number of note-taking sessions varied significantly, ranging from four sessions for some pairs to twenty-eight sessions for others. Specific details regarding the participant period, the number of notes, and emoji usages per pair are provided in Table \ref{table:casestudypair}. During this deployment phase, the logs of their note-taking activities and emoji usage were recorded on our server. 

\subsubsection{Post-deployment Phase}
After a minimum of four weeks of using EmoBridge, participants completed the same set of questions in the questionnaire from the pre-deployment phase (see Table \ref{table:questionnaire}). 
Following the questionnaire, participants were asked to freely share their PNTP experiences with EmoBridge and provide in-depth explanations for their responses to each questionnaire item that they just filled out. Additionally, we shared their usage logs and revisited their notes, prompting them to recall and reflect on the in-class situations where EmoBridge was used for interaction with their PNTP partners. This post-deployment phase session took about one hour and was conducted either in person or remotely. Upon completion of the deployment, participants were compensated 50,000 KRW (equivalent to \$36 USD) for their time and participation.

\subsection{Data Analysis}
For the questionnaire data analysis, given the small sample size and the failure to meet normality test requirements, we utilized the Wilcoxon signed-rank test to compare the responses to each questionnaire item between pre- and post-deployment.

For the qualitative analysis of the interview data, we conducted a thematic analysis of the transcripts using a method that included multiple rounds of both inductive and deductive coding \cite{saldana2021coding}. Initially, we open-coded the characteristics of each pair. Through this process, we identified that class characteristics, relationship dynamics between SWDs and PNTs, and other relevant factors were key in shaping their PNTP experience and EmoBridge usage.
Following this initial step, we sought to understand how these pair-specific characteristics, in conjunction with EmoBridge's features, influenced their note-taking activities while assessing the overall effectiveness of EmoBridge. This approach allowed us to gain deep insights into the nuanced experiences of each SWD-PNT pair, while also identifying broader patterns and themes across the study participants.

\begin{table}[h!]
\centering
\resizebox{\columnwidth}{!}{
\begin{tabular}{cl}
\hline
Number & Question                                                                                                                    \\ \hline
Q1     & Note-taking quality did not meet my expectations.                                                                              \\
Q2     & Note-taking process was challenging.                                                                                    \\
Q3     & I had a good understanding of the contents in the notes.                                                                         \\
Q4     & \begin{tabular}[c]{@{}l@{}}I was able to concentrate well in class while\\taking notes.\end{tabular}                       \\
Q5     & I felt that the partner was relying on me.                                                                                  \\
Q6     & I felt that I could rely on the partner.                                                                                    \\
Q7     & \begin{tabular}[c]{@{}l@{}}I could easily ask the partner about class content\\ I did not understand.\end{tabular}          \\
Q8     & \begin{tabular}[c]{@{}l@{}}I felt as if I was taking note together with partner\\ during the class.\end{tabular}            \\
Q9     & \begin{tabular}[c]{@{}l@{}}I was able to share feedback with the partner \\about the note-taking in real-time.\end{tabular}                 \\
Q10    & \begin{tabular}[c]{@{}l@{}}The process of taking notes together was enjoyable.\end{tabular} \\
Q11    & \begin{tabular}[c]{@{}l@{}} Compared to previous note-taking methods, \\the EmoBridge provided more advantages.\end{tabular}              \\ \hline
\end{tabular}
}
\caption{Questionnaire for evaluating PNTP experience. Each item was rated on a 7-point Likert scale from 1 = Strongly Disagree to 7 = Strongly Agree.}

\Description{This table presents a 7-point Likert scale questionnaire used during the interviews. Each question is rated from 1 to 7, where:
1 = Strongly Disagree,
2 = Disagree,
3 = Slightly Disagree,
4 = Neutral,
5 = Slightly Agree,
6 = Agree,
and 7 = Strongly Agree.
The questions are as follows:
1. The note-taking output does not meet my needs.
2. Process of taking notes was challenging.
3. I think I understand the content of the notes well.
4. I was able to concentrate well in class while taking notes.
5. I felt that the partner was relying on me.
6. I felt that I could rely on the partner.
7. I could easily ask the partner about class content I did not understand.
8. I felt as if I was taking note together with partner during the class.
9. I was able to ask the partner about the note-taking in real-time.
10. The process of taking notes together using the note-taking program was enjoyable.
11. The note-taking program I used is more advanced than other programs.}
\label{table:questionnaire}
\end{table}

\section{Results}
The results of the in-the-wild evaluation study are divided into two sections: Case Study and Effectiveness of EmoBridge. The Case Study section provides a detailed examination of each pair’s characteristics and the specific use cases of EmoBridge. The Effectiveness of EmoBridge section presents an overview of the general effectiveness of EmoBridge, with a focus on its role in improving collaborative partnerships within PNTP.

\subsection{Case Study}

\begin{table*}[h!]
\centering
\resizebox{1\textwidth}{!}{
\begin{tabular}{llllllllll}
\toprule
ID & Gender& Same & Age Diff.& Class Frequency & Participation& \# of & \# of & \# of NT& \# of CC\\
 & (SWD-PNT)& Major& (SWD-PNT)& (per week)& Period & Notes & Emojis & Emojis &Emojis \\
 & & & & & (weeks)& Written& Used& Used&Used\\
\midrule
G1 (SWD-1, PNT-9)& F-F & TRUE & 3 & 1 & 4 & 4 & 29 & 21 & 8\\
G2 (SWD-2, PNT-10)& M-F & TRUE & -1 & 1 (also had meal & 6& 5& 8& 4& 4\\
 & & & & support twice a week)& & & & &\\
G3 (SWD-3, PNT-11)& M-F & TRUE & -4 & 1 & 5 & 5 & 33 & 25 & 8\\
G4 (SWD-4, PNT-12)& F-M & FALSE & -2 & 1 & 5 & 5 & 36 & 24 & 12\\
G5 (SWD-4, PNT-13)& F-F & TRUE & 2 & 4 (2 courses& 7 & 28 & 25 & 20 & 5\\
 & & & & held twice each)& & & & &\\
G6 (SWD-5, PNT-14)& M-M & TRUE & -1 & 1 & 6 & 6 & 4 & 2 & 2\\
G7 (SWD-5, PNT-15)& M-F & TRUE & -2 & 1 & 9 & 7 & 3 & 1 & 2\\
\bottomrule
\end{tabular}
}
\caption{Description of Pairs and Their Utilization of the EmoBridge. The (ID) column represents the IDs assigned to each pair. The (Gender) column indicates the gender of participants (F: female, M: male). The (Same Major) column is marked TRUE if the pair’s majors are the same, and FALSE otherwise. The (Age Diff.) column represents the age difference between the pair. The (Class Frequency) column indicates the number of classes per week, showing how often the pair met weekly. The (Participation Period) column denotes the number of weeks they used EmoBridge. The (\# of Notes Written) column shows the total number of notes taken in EmoBridge. The (\# of Emojis Used) column shows the total number of emojis used during the deployment phase. The (\# of NT Emojis Used) column indicates the total number of note-taking emojis used. The (\# of CC Emojis Used) column shows the total number of chit-chat emojis used.}

\Description{This table provides detailed information about each pair and their utilization of EmoBridge. It includes the following columns: ID, Gender, Major Similarity, Age Difference, Class Frequency, In-the-wild Period, Number of Notes Written, Number of Emojis Used, Number of NT Emojis Used, and Number of CC Emojis Used.
	•	 (ID) indicates the index assigned to participants, with “P” representing a pair.
	•	 (Gender) specifies the gender of participants, where “M” denotes Male and “F” denotes Female.
	•	 (Same Major) is marked TRUE if the pair’s majors are the same, and FALSE otherwise.
	•	 (Age Diff.) represents the age difference between the pair.
	•	 (Class Frequency) indicates the number of classes per week, showing the frequency of the pair’s meetings each week.
	•	 (Participation Period) denotes the duration of the In-the-wild study.
	•	 (# of Notes Written) shows the total number of notes written using EmoBridge.
	•	 (# of Emojis Used) shows the total number of emojis utilized in EmoBridge.
	•	 (# of NT Emojis Used) indicates the number of note-taking emojis used.
	•	 (# of CC Emojis Used) shows the number of chit-chat emojis used.}
\label{table:casestudypair}
\end{table*}

\subsubsection{G1 (SWD-1, PNT-9)}
SWD-1 and PNT-9, from the same department with a three-year age difference, did not know each other before the semester. PNT-9 volunteered for the PNTP to gain insights about department life from colleagues and have a class companion. 

Prior to EmoBridge, PNT-9 took notes during class and shared them with SWD-1 afterward. With EmoBridge, SWD-1 gained the ability to view, understand, and request real-time modifications to the notes. This new dynamic benefited PNT-9 by reducing pressure, as she knew SWD-1 could identify areas for correction or even directly improve the notes.
As their relationship developed, the pair frequently utilized chit-chat emojis for communication during note-taking. SWD-1 used encouraging emojis like 'You Got This' or 'I Like It,' with PNT-9 reciprocating with similar encouraging chit-chat emojis. PNT-9 remarked, ``\textit{I used chit-chat emojis the most when SWD-1 seemed to be struggling with note-taking, as I wanted to casually encourage her}.''

This case illustrated how EmoBridge facilitated real-time collaboration, reduced pressure on the PNT, and fostered a supportive relationship between the pair through emoji-based communication.

\subsubsection{G2 (SWD-2, PNT-10)}
SWD-2 and PNT-10 were already acquainted, sharing the same department and having a small age gap. Their bond was further strengthened by PNT-10's additional role in assisting SWD-2 during meal times, beyond their PNTP participation.

Both participants found EmoBridge enjoyable for facilitating in-class communication. SWD-2 utilized somewhat directive chit-chat emojis, such as 'Stay Awake' when PNT-10 appeared drowsy. SWD-2 also employed 'Difficult' note-taking emojis, which prompted PNT-10 to supplement the notes. Their communication extended beyond emojis to include text within the shared note editor.
PNT-10 remarked, ``\textit{I used many emojis and explained them with text (in the shared note editor in the EmoBridge) when necessary. Emojis allowed quicker, more direct expression, but sometimes text was also helpful}.''

This case showed how EmoBridge enhanced communication between an already familiar pair, allowing for both quick emoji-based interactions and more detailed text explanations. It also demonstrated the platform's flexibility in accommodating different communication modes within the PNTP context.

\subsubsection{G3 (SWD-3, PNT-11)}
SWD-3 and PNT-11, from the same department, had previously participated in PNTP together during the last semester. Despite this, they maintained a somewhat strained relationship, likely influenced by their gender difference and PNT-11 being three years senior to SWD-3. They often sat separately, and occasional class cancellations reduced their meeting frequency.

This pair primarily utilized note-taking emojis rather than chit-chat emojis, with the exception of 'Thank You.' Prior to EmoBridge, they had used Google Docs for real-time note-sharing. Interestingly, after they started using EmoBridge, PNT-11's note-taking style changed. Previously, she usually transcribed everything the lecturer said, but with EmoBridge, she began using more summarized, jotted notes. PNT-11 explained, ``\textit{As I keep an eye on emojis (from the SWD), I kinda stopped myself from fully transcribing the class notes}.'' The awareness that SWD-3 was actively following the lecture and ready to provide feedback on her notes reduced the pressure to capture every word. Despite this shift from full transcription to jotted notes, SWD-3 reported that his satisfaction with her notes remained unchanged. He noted, ``\textit{While I preferred receiving fully transcribed notes as before, I found that summarized notes did not significantly affect my satisfaction (with her notes)}.''

This case demonstrated how EmoBridge could influence established note-taking practices, which highlighted EmoBridge's potential to reduce pressure on PNTs while maintaining the quality and usefulness of notes for SWDs.

\subsubsection{G4 (SWD-4, PNT-12)}
Even though SWD-4 and PNT-12 had different majors and genders, PNT-12's previous knowledge of the course materials motivated him to assist SWD-4 enthusiastically.

PNT-12 made frequent use of the 'Important' note-taking emoji to highlight key points and the 'Focus' emoji to regain SWD-4's attention. He explained, ``\textit{I used the 'Important' and 'Focus' emojis when SWD-4 seemed to lose focus or when critical information was discussed in class. I think it allowed me to easily draw her attention and mark important parts}.'' This emoji-based communication significantly reduced PNT-12's perceived burden and helped him feel fulfilled in his role. He further remarked, ``\textit{Using EmoBridge relieved some burden, as I felt I had fulfilled my role as someone who supports the SWD fully}.'' 

However, there was a mismatch in their interaction styles. While PNT-12 found the emoji features helpful and fulfilling, SWD-4 seemed to perceive his help and motivation as somewhat overwhelming, resulting in her limited use of emojis.

This case illustrated how EmoBridge could facilitate enthusiastic support from PNTs and provide positive emotional reinforcement for their role. However, it also underscored the importance of balancing support with the SWD's comfort level and preferences.

\subsubsection{G5 (SWD-4, PNT-13)}
G5 involved the same SWD as G4 (SWD-4) but with a different PNT (PNT-13). Unlike in G4, SWD-4 and PNT-13 were in the same department and shared the same gender. They met four times a week for two courses, which naturally led to a closer relationship. PNT-13 explained, ``\textit{There were team projects and individual assignments, so I first reached out to SWD-4 to work together. I wanted to receive her help and also wished to become friends}.''

In contrast to her interaction with PNT-12 in G4, SWD-4 utilized emojis more frequently with PNT-13. This difference suggested that the closer relationship in G5 might have reduced the psychological barrier for SWD-4 in using emojis. Additionally, a comparison of note logs revealed that PNT-13 included more contextual information beyond lecture content, unlike PNT-12. Given that SWD-4 occasionally struggled to grasp context due to her hearing impairment, she might have been more attentive to and appreciative of PNT-13's notes with richer contexts.

This case highlighted how the relationship dynamics between SWDs and PNTs could significantly influence the use and effectiveness of tools like EmoBridge. It further demonstrates that note-taking style could impact the comfort level and engagement of SWDs in collaborative note-taking.

\subsubsection{G6 (SWD-5, PNT-14)}
Participants in G6, although from the same department, did not sit together and were not particularly close, which led to a minor misunderstanding between them. 

SWD-5 was highly satisfied with PNT-14's notes and, therefore, did not provide specific feedback. Also, he generally preferred not to use emojis. PNT-14 expressed disappointment at the lack of feedback from SWD-5 but admitted to not using emojis proactively either. He remarked, ``\textit{I felt a bit awkward using chit-chat emojis with him. I would have used them much more if it were with close friends}.''

The situation in G6 underscored the importance of establishing clear communication norms and expectations, especially in pairs that were not particularly close. It also suggested that the effectiveness of emoji-based communication in EmoBridge might be influenced by personal communication preferences and comfort levels, particularly due to perceived social awkwardness or unfamiliarity between users.

\subsubsection{G7 (SWD-5, PNT-15)}
G7 also involved SWD-5 but presented a contrast to G6 in terms of emoji usage. Although the pair in G7 were not close either, they were more proactive in using emojis. This difference primarily stemmed from the class style, which significantly influenced note-taking methods.

In G6's instructor-led lecture class, PNT-14 transcribed notes verbatim, leading to minimal need for feedback. Conversely, in G7's student-led presentation class, PNT-15 needed to summarize the points from different students, which prompted more feedback from SWD-5. SWD-5 commented on this difference, saying, ``\textit{With PNT-14 transcribing sentences verbatim from the instructors, there was little to discuss. In contrast, PNT-15's summaries left more room for conversation}.''

This contrast between G6 and G7 suggested that the effectiveness of emoji-based communication was not solely dependent on the closeness of the relationship, but could be greatly influenced by the context of the class and the nature of the notes being shared.

\subsection{Overall Effects of EmoBridge on PNTP Experience}

Results from the statistical analysis of the questionnaire items were used to assess the overall effects of EmoBridge on their PNTP experiences.

\subsubsection{Effects on Note-taking and learning process}

\begin{figure*}[h!]

  \centering
  \includegraphics[width=\textwidth]{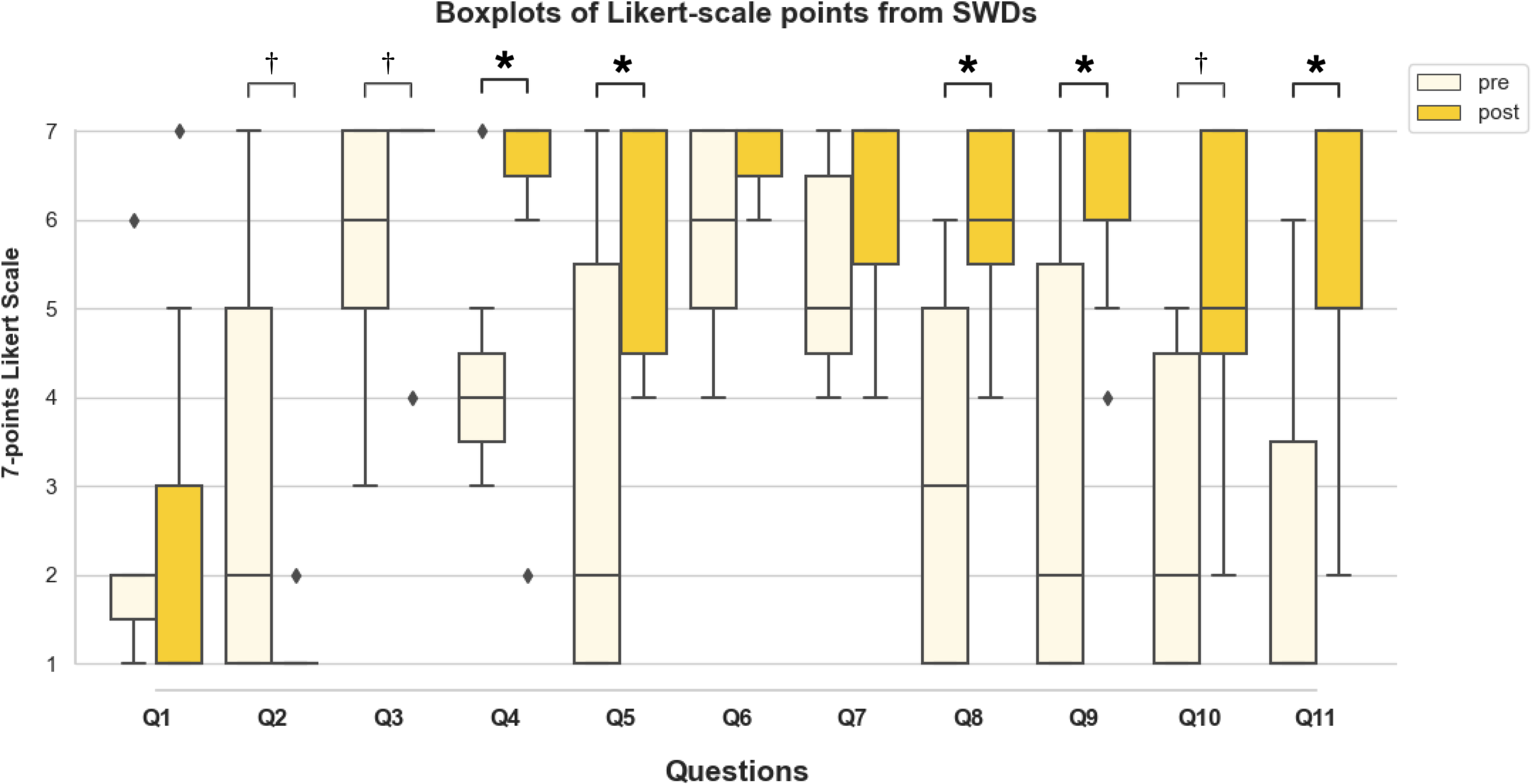}
  \caption{Boxplot Comparison of Pre- and Post-Deployment Questionnaire Responses for SWDs. This figure displays the distribution of responses for each question on a 7-point Likert scale (1 = Strongly Disagree, 7 = Strongly Agree), collected before (pre) and after (post) the deployment. Statistically significant differences (p-value < 0.05) are indicated by asterisks (*), demonstrating the deployment’s impact on the responses. Additionally, questions with p-values between 0.05 and 0.1 are marked with a dagger symbol, indicating marginal significance. Outliers for all questions are indicated by diamonds. For detailed data and statistical analysis, refer to Appendix A.1. The specific questions corresponding to each boxplot are listed in Table \ref{table:questionnaire}. For questions Q2 and Q3, all values except for one are either 1 or 7, resulting in no box being drawn and only a line being shown.}
  
  \Description{The figure titled “Boxplots of Likert-scale points from SWDs” provides a visual comparison of pre- and post-deployment responses from Students With Disabilities (SWDs) on a series of questions. The responses are measured on a 7-point Likert scale, where 1 represents “Strongly Disagree” and 7 represents “Strongly Agree”. The pre-deployment responses are shown in light beige, while the post-deployment responses are depicted in yellow.
  The x-axis of the figure is labeled with the question numbers from Q1 to Q11, representing different survey questions. The y-axis represents the Likert scale points from 1 to 7.
  Each boxplot includes several key features:
	•	The central line within each box represents the median response.
	•	The top and bottom edges of the box represent the interquartile range (IQR), showing where the middle 50
	•	The whiskers extend from the edges of the box to the smallest and largest values within 1.5 times the IQR.
	•	Outliers, which are data points outside the range of the whiskers, are indicated by small diamonds.
  Asterisks (*) above the boxplots for questions Q4, Q5, Q8, Q9, and Q11 indicate statistically significant differences between pre- and post-deployment responses, with a p-value less than 0.05. This signifies that the deployment had a significant impact on the responses for these questions. Additionally, questions with p-values between 0.05 and 0.1 are marked with a dagger symbol, indicating marginal significance. Those questions are questions Q2, Q3, Q10.
  For detailed data and statistical analysis, please refer to Appendix A.1. The specific questions corresponding to each boxplot are listed in Table 3.}
  \label{figure:Boxplot_SWD}
\end{figure*}

\begin{figure*}[h!]
  \centering
  \includegraphics[width=\textwidth]{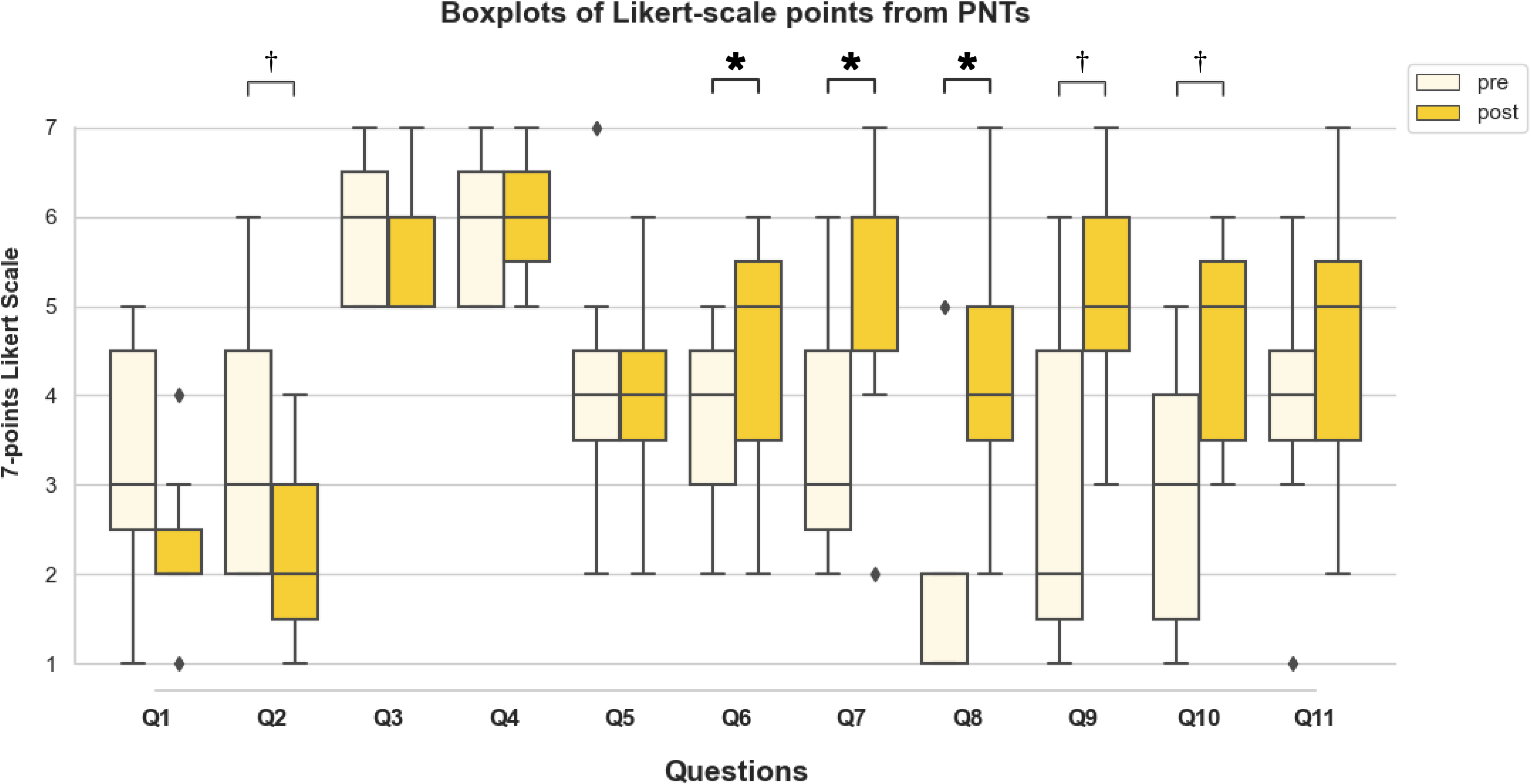}
  \caption{Boxplot Comparison of Pre- and Post-deployment Questionnaire Responses for PNTs. This figure displays the distribution of responses for each question on a 7-point Likert scale (1 = Strongly Disagree, 7 = Strongly Agree), collected before (pre) and after (post) the deployment of EmoBridge. Statistically significant differences (p-value < 0.05) are indicated by asterisks (*). Additionally, items with p-values between 0.05 and 0.1 are marked with a dagger symbol, indicating marginal significance. Outliers for all questions are indicated by diamonds. For detailed data and statistical analysis, refer to Appendix A.2. The specific questions corresponding to each boxplot are listed in Table \ref{table:questionnaire}.}
  
  \Description{The figure titled “Boxplots of Likert-scale points from PNTs” provides a visual comparison of pre- and post-deployment responses from Peer Note Takers (PNTs) on a series of questions. The responses are measured on a 7-point Likert scale, where 1 represents “Strongly Disagree” and 7 represents “Strongly Agree”. The pre-deployment responses are shown in light beige, while the post-deployment responses are depicted in yellow.
  The x-axis of the figure is labeled with the question numbers from Q1 to Q11, representing different survey questions. The y-axis represents the Likert scale points from 1 to 7.
  Each boxplot includes several key features:
	•	The central line within each box represents the median response.
	•	The top and bottom edges of the box represent the interquartile range (IQR), showing where the middle 50
	•	The whiskers extend from the edges of the box to the smallest and largest values within 1.5 times the IQR.
	•	Outliers, which are data points outside the range of the whiskers, are indicated by small diamonds.
  Asterisks (*) above the boxplots for questions Q6, Q7, and Q8 indicate statistically significant differences between pre- and post-deployment responses, with a p-value less than 0.05. This signifies that the deployment had a significant impact on the responses for these questions. Additionally, questions with p-values between 0.05 and 0.1 are marked with a dagger symbol, indicating marginal significance. Those questions are questions Q2, Q9, Q10.
  For detailed data and statistical analysis, please refer to Appendix A.2. The specific questions corresponding to each boxplot are listed in Table 3.}
  \label{figure:Boxplot_PNT}
\end{figure*}

To evaluate EmoBridge, we examined the effects of the EmoBridge  EmoBridge's potential to enhance note quality (Q1: ``\textit{Note-taking quality did not meet my expectations}.'' ),  the challenge of the note-taking process (Q2: ``\textit{Note-taking process was challenging}.'' ), and note content understanding (Q3: ``\textit{I had a good understanding of the contents in the notes}.'' ). Although responses indicated slight improvements across these areas, they did not achieve statistical significance for both SWD and PNT participants (see Fig. \ref{figure:Boxplot_SWD} and Fig. \ref{figure:Boxplot_PNT}). However, there was a trend towards improvement in addressing note-taking challenges for both groups (Q2). SWDs reported a reduction in challenges (Pre-deployment: Median=2, IQR=3; Post-deployment: Median=1, IQR=0; p-value=0.06), as did PNTs (Pre-deployment: Median=3, IQR=2.5; Post-deployment: Median=2, IQR=1.5; p-value=0.06). While not statistically significant, these trends suggested that EmoBridge might have eased some of the difficulties associated with the note-taking process for both SWDs and PNTs.

Regarding class concentration, we observed significant improvements for SWDs but not for PNTs. In the pre-deployment phase, responses to Q4 (``\textit{I was able to concentrate well in class while taking notes}.'') were neutral (Median=4, IQR=1). However, post-deployment results showed a significant improvement for SWDs, who reported significantly better concentration during the note-taking process (Median=7, IQR=0.5, p-value=0.04). Analysis of the post-deployment interviews also echoed that many SWD participants felt significantly enhanced participation in class. SWD-5 explained, ``\textit{Previously when I missed the point during class, I just had to let it go and wait until the end of the class to check notes. But now I don't miss the point as I can see every note taken in real-time. This makes me feel less left out.}'' Similarly, SWD-4 mentioned, ``\textit{Before, it felt like I was just passively listening to the class, but now with EmoBridge, I can express when I'm hard to keep up. It makes me feel more like a part of the class.}''
In contrast, PNTs did not show significant improvements in this area (p-value=0.70). This suggested that EmoBridge's real-time sharing features might have particularly enhanced SWDs' ability to engage with and focus on class content, while not significantly affecting PNTs' concentration levels.

The enjoyment of the note-taking process (Q10: ``\textit{The process of taking notes together using the note-taking program was enjoyable}.'') showed a positive trend for both SWDs and PNTs after using EmoBridge. In the pre-deployment phase, SWDs generally reported low enjoyment or non-participation in collaborative note-taking, with a median score of 2 (IQR=3.5). Post-deployment results, however, indicated a trend toward more positive feelings about the collaborative note-taking process (Median=5, IQR=2.5, p-value=0.08). Similarly, PNTs reported a trend toward increased enjoyment. Their pre-deployment ratings were relatively low (Median=3, IQR=2.5), but post-deployment scores showed some improvement (Median=5, IQR=2, p-value=0.06). While these improvements did not reach statistical significance, the consistent trend across both groups indicated that EmoBridge might have enhanced the overall experience of collaborative note-taking, making it more enjoyable for both SWDs and PNTs. 

Finally, there were concerns that EmoBridge's features might not be perceived as an advancement over existing note-taking methods. However, the results showed significant improvements, particularly for SWDs. In the pre-deployment assessment, when first introduced to EmoBridge, SWDs strongly disagreed with Q11 (``\textit{Compared to previous note-taking methods, EmoBridge provided more advantages}.''), based on their initial impressions (Median=1, IQR=2.5).
In contrast, post-deployment responses to Q11 indicated a statistically significant enhancement, as shown in Fig. \ref{figure:Boxplot_SWD}. SWDs rated EmoBridge as considerably more advantageous compared to previous methods (Median=5, IQR=2, p-value=0.041). This significant shift in SWD participants' assessments from pre- to post-deployment phases underscored the potential value of EmoBridge's collaborative features in addressing specific needs of SWDs that were not met by previously used note-taking methods.

\subsubsection{Effects on In-Class Communication and Collaborative Partnership}
In terms of in-class communication regarding note-taking and lecture content, PNT participants initially found it challenging to ask SWDs about unclear content. Pre-deployment responses to Q7 (``\textit{I could easily ask the partner about class content I did not understand}.'') were low for PNTs (Median=3, IQR=2) compared to SWDs (Median=5, IQR=2). This indicated a pre-existing difficulty for PNTs in soliciting feedback from SWDs. However, post-deployment results showed a significant improvement, with PNTs reporting increased ease in asking SWDs about unclear content (Median=6, IQR=1.5, p-value=0.039).

Furthermore, both SWD and PNT participants initially perceived that sharing real-time feedback during class was challenging. Pre-deployment responses to Q9 (``\textit{I was able to share feedback with the partner about the note-taking in real-time}.'') were low for both groups (SWD: Median=2, IQR=4.5; PNT: Median=2, IQR=3). However, post-deployment data revealed significant improvements. All SWDs reported high availability of feedback in note-taking (Median=7, IQR=1, p-value=0.042), while PNTs showed a trend towards improved ability to share feedback with their SWD partners (Median=5, IQR=1.5, p-value=0.05). These findings suggested that EmoBridge significantly enhanced in-class communication between SWDs and PNTs by facilitating an easier exchange of questions and feedback about lecture content and note-taking in real-time. 

The most notable change observed was the potential enhancement of collaborative partnerships between SWDs and PNTs after using EmoBridge. This improvement was evident in both SWDs' and PNTs' responses to Q8 (``\textit{I felt as if I was taking notes together with my partner during the class}.''). In the pre-deployment phase, many SWDs generally disagreed with this statement (Median=3, IQR=4). However, post-deployment results showed a significant shift, with all SWD participants reporting collaborative note-taking with their PNTs. The median response changed from 3 (Slightly Disagree) to 6 (Agree) (Median=6, IQR=1.5, p-value=0.04). Similarly, PNTs initially reported a low sense of collaboration with SWDs (Median=2, IQR=1). Post-deployment, their perception significantly changed, indicating a stronger feeling of collaborative note-taking (Median=4, IQR=1.5, p-value=0.043). These statistically significant improvements for both SWDs and PNTs suggested that EmoBridge effectively fostered a sense of collaboration between partners. 

In particular, the post-deployment interview data also revealed that PNTs felt they could depend more on SWDs due to two main factors: SWDs' direct involvement in note correction and the in-class communication channels provided. PNT-9 explained how SWDs' direct corrections enhanced collaboration: ``\textit{There might have been something I missed, but my partner would write it down in the middle of the session. SWD being able to edit the note directly, I felt more like we were working together}.'' Additionally, having convenient in-class communication channels reduced psychological stress for PNTs. PNT-11 noted, ``\textit{Having a communication channel makes me feel reassured. With emojis, SWD can quickly give me feedback on my notes, which has been helpful. If SWD-3 isn't using note emojis, maybe it means I'm doing okay}.''

Furthermore, the findings revealed a potential for mutual support and interdependence between SWD and PNT participants facilitated by EmoBridge. Pre-deployment data showed that SWD participants did not strongly feel that their PNTs were reliant on them (Q5: ``\textit{I felt that the partner was relying on me}'', Median=2, IQR=4). However, post-deployment responses indicated a significant shift, with SWD participants reporting a strong sense of reliance from their PNTs (Median=7, IQR=2.5, p-value=0.016). Correspondingly, as SWD paticipants perceived increased reliance from PNTs, PNT participants themselves reported an increased reliance on SWDs, which alleviated their mental burden of sole responsibility for note-taking. In the pre-deployment phase, PNTs expressed a neutral view on Q6 (``\textit{I felt that I could rely on the partner}'', Median=4, IQR=1.5). Post-deployment results, however, indicated a significantly enhanced sense of reliance on SWD participants (Median=5, IQR=2, p-value=0.025). These findings suggested that EmoBridge might have fostered a more balanced and interdependent relationship between SWDs and PNTs, where SWDs felt more valued and contributed, while PNTs experienced reduced pressure through shared responsibility.

The post-deployment interview data further demonstrated the enhancement of collaborative partnerships and interdependence through the use of EmoBridge. SWD-2 shared a revealing episode: ``\textit{I remember when PNT-10 was so sleepy that she said she couldn't take notes properly. Therefore, I sent her a bunch of emojis to keep her awake. It was fun, and I felt she began to rely on me}.'' This anecdote illustrated how EmoBridge facilitated a more supportive and interdependent relationship between SWDs and PNTs.

\section{Discussion}
In this section, we summarize our findings and discuss how collaborative note-taking can be integrated into PNTPs and foster interdependence. We then propose design guidelines for collaborative note-taking tools to improve the efficacy of PNTPs and enhance student accessibility and disability support in higher education. 

\subsection{Integrating Collaborative Note-Taking into Peer Note-Taking Program}
Our formative study reaffirmed that communication within PNTPs tends to be unidirectional \cite{Kim2011, ChoSon2013}. We also identified a mutual desire among participants to exchange real-time feedback, increase in-class interaction, and enhance mutual engagement in note-taking activities. These findings informed the design requirements for EmoBridge, a tool we developed to facilitate collaborative note-taking for PNTP contexts.

To explore EmoBridge's potential, we conducted an in-the-wild deployment study with seven SWD-PNT pairs, tasking each pair using EmoBridge to take notes during their classes. This approach allowed us to gather firsthand accounts of the EmopBridge's effectiveness in real classroom settings.

Our findings demonstrated the feasibility of collaborative note-taking in PNTPs, indicating its potential to enhance the effectiveness of PNTPs. Notably, SWD participants reported feeling more involved in their classes, suggesting that collaborative note-taking tools like EmoBridge could foster greater engagement among SWDs in PNTP contexts. This aligns with the well-known benefits of collaborative note-taking in general educational settings \cite{zyto_successful_2012}. Moreover, the ability to see others' contributions to the notes and share responsibility appeared to reduce stress for PNTs. This suggests that collaborative note-taking could not only facilitate better class experiences for SWDs but also alleviate the workload and emotional burden on PNTs. This dual benefit echoes the advantages of collaborative note-taking observed in other educational contexts \cite{costley2021collaborative}. By fostering greater engagement for SWDs and reducing the burden on PNTs, collaborative note-taking tools could significantly enhance the effectiveness and satisfaction of PNTP participants, potentially leading to improved academic outcomes and a more positive educational experience overall.

\subsection{Collaborative Note-Taking for Growing Interdependence between Students with Disabilities and their Peer Note-takers}

Our findings revealed that collaborative note-taking facilitated by EmoBridge evoked a sense of interdependence, a framework that emphasizes collaborative access and the contributions of people with disabilities in a task \cite{bennett2018interdependence}. This observation was evidenced by three key outcomes. First, both parties perceived the PNTP as more of a collaboration, shifting from a unidirectional support model to a more interactive partnership. Second, SWD participants reported experiencing a sense of self-efficacy by actively contributing to the notes. Third, PNT participants developed a reliance on SWDs, recognizing their valuable input in the note-taking process.

These outcomes suggest that EmoBridge fostered a more balanced and mutually beneficial relationship between SWDs and PNTs. By enabling SWDs to contribute actively and allowing PNTs to rely on these contributions, collaborative note-taking tools like EmoBridge have potential to create a more interdependent dynamic within the PNTP. This shift towards interdependence could potentially lead to more effective note-taking, increased engagement from SWDs, and a more satisfying experience for both parties involved in the PNTP.

\subsection{Collaborative Note-taking Tool Design Guideline}

Based on the findings of our study and insights gained while developing and evaluating EmoBridge, we provide the following design guidelines for integrating collaborative note-taking tools into PNTP contexts:

\subsubsection{Enable Real-Time Note-Taking Screen Sharing}
Emobridge enables real-time sharing of notes and cursor positions among users. This feature gave participants a sense of attending classes together, aka social presence, the sense of being together further,  even though they were physically together in the class room, because of the procedure variances of the PNTPs or closeness between them their sense of being together was relatively low. , fostering a sense of collaboration and a greater reliance on each other. 
The ability for SWDs to view and edit notes in real-time could enhance their understanding and engagement, especially beneficial for those with hearing disabilities \cite{cerva2012real}. This interaction not only helped SWDs participate actively in the note-taking process, fostering interdependence but also eased the mental load on PNTs who can improve note quality through immediate feedback. Thus, a real-time screen-sharing feature that allows both SWDs and PNTs to contribute should be equipped.

\subsubsection{Enable Concise but Flexible In-class Communication Methods}
In developing EmoBridge, we chose emojis for in-class communication due to their convenience, ease of use, and quick input. As such, we believe that methods facilitating in-class communication between SWDs and PNTs should minimize disruption to attention and note-taking themselves.

In doing so, the interaction method for in-class communication should be quick and easy to use. Particularly, the input mechanism for feedback on shared note areas needs to be implemented with consideration for the various disabilities that SWDs might have, which could prevent them from using the note-taking interface freely. Even for SWDs without such limitations, the system should enable efficient feedback management during note-taking. 

At the same time, it is crucial to acknowledge that individual preferences for communication modes may vary. Some participants in our study felt that using emojis with individuals they were not personally close to was socially awkward. This highlights the need for flexibility in communication options. Beyond emojis, incorporating shortcuts or symbols that are easily inputted via keyboard or mouse could facilitate communication while potentially reducing social discomfort. These alternative input methods could provide a more neutral or formal means of interaction, which might be preferable in certain PNTP relationships or academic settings. For instance, predefined text shortcuts or customizable symbols could offer quick, efficient communication without the perceived informality of emojis. 

\subsubsection{Enable Both Note-Driven and Social In-class Communication}
EmoBridge provided two sets of emojis: note-taking emojis for sharing feedback on notes and chit-chat emojis for social communication. Our in-the-wild evaluation findings showed that this categorization could be beneficial in supporting both the practical aspects of collaborative note-taking and the social dimension of the PNTP relationship.

For supporting effective note-driven communication, it is crucial to have clear pre-set input messages allowing precise feedback on particular aspects of the notes. The note-taking emojis in the EmoBridge served this purpose, enabling quick and targeted communication about the content being recorded. This approach could enhance collaborative learning by facilitating an efficient feedback loop between SWDs and PNTs.

In supporting social communication, chit-chat emojis allow for interactions in the broader lecture context. As such, this distinction between note-related feedback and more general social interaction can enhance the overall learning experience and foster collaborative relationships between SWDs and PNTs by providing a way for partners to express gratitude or share brief personal reactions without disrupting the primary task of note-taking.

By providing these dual channels of communication, collaborative note-taking tools can support both the academic goals of the PNTP and the development of a supportive partnership between SWDs and PNTs. This balanced approach has the potential to create a more engaging and satisfying experience for both parties, enhancing not only the quality of notes but also the sense of shared participation in the learning process.

\subsubsection{Both Relationship Dynamics and Class Contexts Need to be Considered}
Our case study revealed that relationship dynamics between SWD and PNT and class contexts significantly influenced the usage and satisfaction with EmoBridge. This finding underscores the importance of tailoring collaborative note-taking tools to each pair's specific characteristics. 

\paragraph{The Degree of Closeness in Relationships}
PNTs and SWDs share a unique relationship in their classes, where one assists the other. Our research found that neither party prefers a purely transactional relationship, while the desired closeness varied significantly. Therefore, designing for the preferred interaction mode is crucial. For example, for closer relationships, informal communication like emojis and memes, as used in EmoBridge, can be effective. However, if a task-focused relationship is preferred, more formal communication methods might be necessary \cite{oluga2013exploration}.

\paragraph{Instructional style of the class}
The instructional style of the class in which PNTP is employed significantly influences design considerations, particularly the distinction between instructor-led and student-centered teaching. Usually, the notes for instructor-led classes require detailed note-taking, while student-centered settings need summaries of discussions \cite{shadiev2017investigating}. By considering these aspects in the design and implementation of collaborative note-taking tools, it may be possible to create more versatile and effective solutions that can adapt to varying characteristics of classes.

\subsection{Limitation and Future Work}
Our study had several limitations that provide opportunities for future research. 

First, we focused on students with disabilities who could read the notes using their vision because our formative study results informed us that students with visual impairments required a different approach compared to students with other types of disabilities. However, future research should include students with visual impairments as well. Adapting existing systems or developing methods tailored to the challenges of visually impaired students could provide a more comprehensive solution.

Second, while our evaluation studies showed EmoBridge's effectiveness, there is more room for improvement. Particularly, the integration of EmoBridge with existing Learning Management Systems (LMS) may significantly enhance its utility and user experience. By linking EmoBridge with LMS, SWDs and PNTs could seamlessly access course materials, assignments, and notes all in one platform, simplifying resource management and possibly increasing student engagement. Further, incorporating speech-to-text (STT) and automated summaries using large language models (LLMs) could be viable directions. Therefore, future studies will need to examine how such AI-based tools could leverage PNTPs.

Finally, while our formative and evaluation studies yield valuable insights, they have limitations due to the small sample size and the fact that all participants were from South Korea. Therefore, future studies will need to expand the participant pool to encompass a more diverse and larger sample.

\section{Conclusion}
Students with disabilities (SWDs) often face challenges in note-taking during lectures, prompting many higher education institutions to implement peer note-taking programs (PNTPs). In these programs, peer note-takers (PNTs) assist SWDs with lecture notes. To gain deeper insights into the experiences of both groups, we conducted semi-structured interviews with eight SWDs and eight PNTs. Our findings revealed predominantly unidirectional interaction between SWDs and PNTs, highlighting specific needs and challenges. In response, we developed EmoBridge, a collaborative note-taking platform facilitating real-time interaction between PNT-SWD pairs using emojis. We evaluated EmoBridge through an in-the-wild study with seven PNT-SWD pairs. 

Our findings suggested that EmoBridge fostered a more balanced and interdependent relationship between SWDs and PNTs, where SWDs felt more valued and contributed, while PNTs experienced reduced pressure through shared responsibility. Such emerging interdependence between SWDs and PNTs suggests mutual benefits that collaborative note-taking could offer to participants of PNTPs. Based on the findings of our study and insights gained while developing and evaluating EmoBridge, we provide the design guidelines for collaborative note-taking systems aimed at enhancing PNTPs and creating more effective and inclusive educational experiences for SWDs. We hope this paper contributes to the ongoing effort to improve accessibility and support in higher education for students with disabilities.

\begin{acks}
We are deeply grateful for our participants and reviewers who significantly contributed to this work. This research was supported by the New Faculty Startup Fund from Seoul National University (\#200-20230022) and the Undergraduate Research Learner (URL) program of Information Science and Culture Studies at Seoul National University.
\end{acks}

\bibliographystyle{ACM-Reference-Format}
\bibliography{ref}

\newpage
\begin{appendices}
\onecolumn
\section{Appendix}

\subsection{Table of Comparative Analysis of Pre- and Post-deployment Data from SWDs}

\begin{table}[hbt!]
\resizebox{\textwidth}{!}{
\begin{tabular}{|cc|c|c|c|c|c|c|c|c|c|c|c|}
\hline
\multicolumn{2}{|c|}{}                                                                                              & \textbf{Q1}    & \textbf{Q2}    & \textbf{Q3}    & \textbf{Q4}              & \textbf{Q5}              & \textbf{Q6}    & \textbf{Q7}    & \textbf{Q8}              & \textbf{Q9}              & \textbf{Q10}   & \textbf{Q11}             \\ \hline
\multicolumn{1}{|c|}{\multirow{2}{*}{\textbf{\begin{tabular}[c]{@{}c@{}}Pre-\\ deployment\end{tabular}}}}  & Median & \textit{2}     & \textit{2}     & \textit{6}     & \textit{4}               & \textit{2}               & \textit{6}     & \textit{5}     & \textit{3}               & \textit{2}               & \textit{2}     & \textit{1}               \\ \cline{2-13} 
\multicolumn{1}{|c|}{}                                                                                     & IQR    & \textit{0.5}   & \textit{3}     & \textit{2}     & \textit{1}               & \textit{4}               & \textit{2}     & \textit{2}     & \textit{4}               & \textit{4.5}             & \textit{3.5}   & \textit{2.5}             \\ \hline
\multicolumn{1}{|c|}{\multirow{2}{*}{\textbf{\begin{tabular}[c]{@{}c@{}}Post-\\ deployment\end{tabular}}}} & Median & \textit{1}     & \textit{1}     & \textit{7}     & \textit{7}               & \textit{7}               & \textit{7}     & \textit{7}     & \textit{6}               & \textit{7}               & \textit{5}     & \textit{5}               \\ \cline{2-13} 
\multicolumn{1}{|c|}{}                                                                                     & IQR    & \textit{2}     & \textit{0}     & \textit{0}     & \textit{0.5}             & \textit{2.5}             & \textit{0.5}   & \textit{1.5}   & \textit{1.5}             & \textit{1}               & \textit{2.5}   & \textit{2}               \\ \hline
\multicolumn{2}{|c|}{p-value}                                                                                       & \textit{0.480} & \textit{0.066\textdagger} & \textit{0.059\textdagger} & \textit{\textbf{0.045*}} & \textit{\textbf{0.016*}} & \textit{0.108} & \textit{0.276} & \textit{\textbf{0.047*}} & \textit{\textbf{0.042*}} & \textit{0.083\textdagger} & \textit{\textbf{0.041*}} \\ \hline
\end{tabular}
}
\caption{Table of Comparative Analysis of Pre- and Post-deployment Data from SWDs. This table presents the statistical outcomes for SWDs from the Likert-scale questionnaire, analyzed using the Wilcoxon signed-rank test. Medians and interquartile ranges (IQR) are reported for both pre- and post-deployment phases. The p-values indicate the significance of changes between these two phases, with an asterisk marking the questions that showed a statistically significant difference, using a threshold of 0.05. Additionally, questions with p-values between 0.05 and 0.1 are marked with a dagger symbol, indicating marginal significance.}

\Description{This table presents the statistical analysis outcomes from the questionnaire for evaluating SWD's PNTP experiences before and after utilizing the EmoBridge, analyzed using the Wilcoxon signed-rank test. Medians and interquartile ranges (IQR) are reported for both pre- and post-deployment phases. The p-values indicate the significance of changes between these two phases, with an asterisk marking the questions that showed a statistically significant difference, using a threshold of 0.05.
The table is divided into two main sections: Pre-deployment and Post-deployment, each reporting the Median and Interquartile Range (IQR) for responses to 11 questions (Q1 to Q11). The pre-deployment median responses range from 1 to 6, while the IQR varies between 0.5 and 4. In the post-deployment phase, the median responses are mostly higher, ranging from 5 to 7, except for Q1 and Q2, which have a median of 1. The IQR in the post-deployment section shows less variability, ranging from 0 to 2.5.
The p-values listed in the bottom row assess the statistical significance of changes from pre- to post-deployment for each question. They range from 0.016 to 0.480. Statistically significant changes, marked with an asterisk, are noted in questions Q3, Q4, Q5, Q8, Q9, and Q11, indicating p-values less than 0.05. Additionally, questions with p-values between 0.05 and 0.1 are marked with a dagger symbol, indicating marginal significance. Those are questions Q2, Q3, Q10. This table is essential for understanding the changes in responses from before and after the deployment, especially noting the significant areas where the deploymental intervention had an impact as per the Wilcoxon signed-rank test results.}
\label{table:statisticsofswds}
\end{table}

\subsection{Table of Comparative Analysis of Pre- and Post-deployment Data from PNTs}

\begin{table}[h]
\resizebox{\textwidth}{!}{
\begin{tabular}{|cc|c|c|c|c|c|c|c|c|c|c|c|}
\hline
\multicolumn{2}{|c|}{}                                                                                              & \textbf{Q1}    & \textbf{Q2}    & \textbf{Q3}    & \textbf{Q4}    & \textbf{Q5}    & \textbf{Q6}              & \textbf{Q7}              & \textbf{Q8}              & \textbf{Q9}    & \textbf{Q10}   & \textbf{Q11}            \\ \hline
\multicolumn{1}{|c|}{\multirow{2}{*}{\textbf{\begin{tabular}[c]{@{}c@{}}Pre-\\ deployment\end{tabular}}}}  & Median & \textit{3}     & \textit{3}     & \textit{6}     & \textit{6}     & \textit{4}     & \textit{4}               & \textit{3}               & \textit{2}               & \textit{2}     & \textit{3}     & \textit{4}              \\ \cline{2-13} 
\multicolumn{1}{|c|}{}                                                                                     & IQR    & \textit{2}     & \textit{2.5}   & \textit{1.5}   & \textit{1.5}   & \textit{1}     & \textit{1.5}             & \textit{2}               & \textit{1}               & \textit{3}     & \textit{2.5}   & \textit{1}              \\ \hline
\multicolumn{1}{|c|}{\multirow{2}{*}{\textbf{\begin{tabular}[c]{@{}c@{}}Post-\\ deployment\end{tabular}}}} & Median & \textit{2}     & \textit{2}     & \textit{6}     & \textit{6}     & \textit{4}     & \textit{5}               & \textit{6}               & \textit{4}               & \textit{5}     & \textit{5}     & \textit{5}              \\ \cline{2-13} 
\multicolumn{1}{|c|}{}                                                                                     & IQR    & \textit{0.5}   & \textit{1.5}   & \textit{1}     & \textit{1}     & \textit{1}     & \textit{2}               & \textit{1.5}             & \textit{1.5}             & \textit{1.5}   & \textit{2}     & \textit{1.5}            \\ \hline
\multicolumn{2}{|c|}{p-value}                                                                                       & \textit{0.219} & \textit{0.066\textdagger} & \textit{0.564} & \textit{0.705} & \textit{0.705} & \textit{\textbf{0.025*}} & \textit{\textbf{0.039*}} & \textit{\textbf{0.043*}} & \textit{0.058\textdagger} & \textit{0.063\textdagger} & \textit{0.157} \\ \hline
\end{tabular}
}
\caption{Table of Comparative Analysis of Pre- and Post-deployment Data from PNTs. This table presents the statistical outcomes for PNTs from the Likert-scale questionnaire, analyzed using the Wilcoxon signed-rank test. Medians and interquartile ranges (IQR) are reported for both pre- and post-deployment phases. The p-values indicate the significance of changes between these two phases, with an asterisk marking the questions that showed a statistically significant difference, using a threshold of 0.05. Additionally, questions with p-values between 0.05 and 0.1 are marked with a dagger symbol, indicating marginal significance.}

\Description{This table presents the statistical outcomes for PNTs from the Likert-scale questionnaire, analyzed using the Wilcoxon signed-rank test. Medians and interquartile ranges (IQR) are reported for both pre- and post-deployment phases. The p-values indicate the significance of changes between these two phases, with an asterisk marking the questions that showed a statistically significant difference, using a threshold of 0.05.
The table presents pre- and post-deployment statistics for a questionnaire with 11 questions (Q1 to Q11). Each question’s median and interquartile range (IQR) are given for both the pre-deployment and post-deployment periods.
For the pre-deployment phase, median scores range from 3 to 6, while IQRs are between 1 and 2.5. In the post-deployment phase, median scores range from 2 to 6, with most median scores being 4 or 5. The IQRs in the post-deployment phase have decreased overall, with values between 0.5 and 2, indicating less variability in responses.
The p-values at the bottom indicate the probability that the changes observed could be due to chance. Values marked with an asterisk (*) are those with p-values less than 0.05, suggesting statistically significant changes. Questions Q6, Q7, and Q8 show significant changes with p-values of 0.025, 0.039, and 0.043, respectively. Additionally, questions with p-values between 0.05 and 0.1 are marked with a dagger symbol, indicating marginal significance. Those are questions Q2, Q9, Q10. 
This statistical overview provides a clear understanding of the changes in participant responses from before to after the deployment, highlighting specific areas with significant changes based on the Wilcoxon signed-rank test.}
\label{table:statisticsofpnts}

\end{table}

\end{appendices}

\end{document}